\documentclass[12pt,preprint]{aastex}
\usepackage{epsfig}
\usepackage{natbib}
\usepackage{graphicx}
\usepackage{slashbox}
\usepackage{multirow}
\usepackage{lscape}
\usepackage{mathrsfs,amssymb}
\usepackage{amsmath}
\newcommand       \Angstrom     {\,{\rm \AA}}
\newcommand       \AngstromAL {{\rm \AA}}

\newcommand       \cm           {\,{\rm cm}}

\newcommand       \kpc          {\,{\rm kpc}}

\newcommand       \nH           {n_{\rm H}}
\newcommand       \simlt        {\lesssim}
\newcommand       \simgt        {\gtrsim}

\newcommand       \mum          {\,{\rm \mu m}}

\newcommand       \simali       {\sim\,}
\newcommand       \magni    {\,{\rm mag}}

\newcommand  \fdib      {f_{\scriptscriptstyle\rm DIB}}
\newcommand  \NDIB  {N_{\scriptscriptstyle\rm DIB}}
\newcommand  \WDIB  {W_{\scriptscriptstyle\rm DIB}}
\newcommand  \WDIBp  {W^{\prime}_{\scriptscriptstyle\rm DIB}}

\newcommand       \RV           {{R_V}}
\newcommand       \AV           {{A_V}}

\newcommand       \Alambda  {A_\lambda}
\newcommand       \lambdaFUV {\lambda_{\scriptscriptstyle\rm FUV}}
\newcommand       \lambdaVIS {\lambda_{\scriptscriptstyle\rm VIS}}

\newcommand       \AFUVint {A_{\scriptscriptstyle\rm FUV}^{\rm int}}
\newcommand  \AJFUVint {\left(AJ\right)_{\scriptscriptstyle\rm FUV}^{\rm int}}
\newcommand  \AJFUVatten {\left(AJ\right)_{\scriptscriptstyle\rm FUV}^{\rm atten}}
\newcommand       \AFUV    {A_{\scriptscriptstyle\rm FUV}}
\newcommand       \NFUV    {N_{\scriptscriptstyle\rm FUV}}

\newcommand       \JISRF    {J_\lambda^{\scriptscriptstyle\rm ISRF}}
\newcommand       \AVISint {A_{\scriptscriptstyle\rm VIS}^{\rm int}}

\newcommand       \Abumpint {A_{\scriptscriptstyle\rm bump}^{\rm int}}

\newcommand       \mA     {\,{\rm m\AA}}
\def    \beq    {\begin{equation}}
\def    \eeq    {\end{equation}}
\def    \beqa   {\begin{eqnarray}}
\def    \eeqa   {\end{eqnarray}}
\shorttitle{DIBs and UV Extinction}
\title{
 \vspace*{-2.0em}
{\normalsize\rm accepted for publication 
in {\it The Astrophysical Journal}}\\
 \vspace*{1.0em}
Diffuse Interstellar Bands and the Ultraviolet Extinction Curves: 
The Missing Link Revisited
\\{\small DRAFT: \today ~~}
}

\author{F.Y. Xiang\altaffilmark{1,2}, Aigen Li\altaffilmark{2} 
                                and J.X. Zhong\altaffilmark{1,2}}
\altaffiltext{1} {Department of Physics, Xiangtan University, 
                  411105 Xiangtan, Hunan Province, China;
                  {\sf fyxiang, jxzhong@xtu.edu.cn}}
\altaffiltext{2} {Department of Physics and Astronomy,
                  University of Missouri,
                  Columbia, MO 65211, USA;
                  {\sf lia@missouri.edu}}

\begin{document}

\begin{abstract}
%
%
A large number of interstellar absorption features 
at $\simali$4000$\Angstrom$--1.8$\mum$,
known as the  ``diffuse interstellar bands'' (DIBs),
remains unidentified. 
Most recent works relate 
them to large polycyclic 
aromatic hydrocarbon (PAH) molecules 
or ultrasmall carbonaceous grains
which are also thought
to be responsible for the 2175$\Angstrom$ 
extinction bump and/or the far ultraviolet (UV) 
extinction rise at $\lambda^{-1}>5.9\mum^{-1}$.
%
Therefore, one might expect some relation 
between the UV extinction and DIBs.
Such a relationship, if established, could put 
important constraints on the carrier of DIBs.
Over the past four decades, 
whether DIBs are related to
the shape of the UV extinction curves has been 
extensively investigated. However, the results 
are often inconsistent, 
partly due to the inconsistencies 
in characterizing the UV extinction. 
Here we re-examine the connection between 
the UV extinction curve and DIBs.
We compile the extinction curves and
the equivalent widths of 40 DIBs
along 97 slightlines.
We decompose the extinction curve into 
three Drude-like functions composed of 
the visible/near-infrared component, 
the 2175$\Angstrom$ bump, 
and the far-UV extinction 
at $\lambda^{-1}>5.9\mum^{-1}$.
We argue that the wavelength-integrated 
far-UV extinction derived from this 
decomposition technique 
best measures the strength of
the far-UV extinction. 
No correlation is found 
between the far-UV extinction 
and most ($\simali$90\%) of the DIBs. 
We have also shown that 
the color excess $E(1300-1700)$,
the extinction difference at 1300$\Angstrom$
and 1700$\Angstrom$ often used to measure
the strength of the far-UV extinction, 
does not correlate with DIBs.  
Finally, we confirm the earlier findings of 
no correlation between the 2175$\Angstrom$ bump 
and DIBs or between the 2175$\Angstrom$ bump and
the far-UV extinction. 

\end{abstract}

\section{Introduction\label{sec:intro}}
The year of 2016 marks the 82nd anniversary 
of the first recognition of the interstellar 
nature of the so-called 
``diffuse interstellar bands'' 
(DIBs; Merrill et al.\ 1934).
The DIBs are a set of several hundred 
absorption features spanning the wavelength 
range of the near ultraviolet (UV) 
at $\lambda\simgt 4000\Angstrom$ 	
(e.g., see York et al.\ 2006, 
Gredel et al.\ 2011, Salama et al.\ 2011, 
Bhatt \& Cami 2015)
to the near infrared (IR) 
at $\lambda\simlt 1.8\mum$
(e.g., see Joblin et al.\ 1990, 
Geballe et al.\ 2011, Cox et al.\ 2014, 
Zasowski et al.\ 2015, Hamano et al.\ 2015),
with most of the bands falling
in the visible wavelength range.
%
%
%
%
The DIBs were originally discovered in 1919 
by Herger (1922) and their interstellar nature
was established based on their stationary nature
as observed toward spectroscopic binaries.
Having full widths at half-maximum (FWHMs) 
ranging from $\simali$0.4 to $\simali$40$\Angstrom$,
they are conspicuously broader than the sharp 
absorption lines of simple atomic and diatomic 
molecules such as CH and CN (see Cox 2011).

Almost every article on DIBs starts by stating that, 
``... since the discovery that DIBs are interstellar'', 
as D\'esert et al.\ (1995) put it,
``the nature of their carriers is still unknown''. 
Unfortunately, this is still true. 
%
Although numerous atomic, molecular, and solid-state 
carriers have been proposed (see Sarre 2006),
so far the vast majority of the DIBs 
remain unidentified and
the origin of the DIBs still remains enigmatic 
(see Cami \& Cox 2014).
Very recently,  Campbell et al.\ (2015, 2016) 
and Walker et al.\ (2015) measured the gas-phase
spectrum of C$_{60}^{+}$ and found that the spectral 
characteristics of gas-phase C$_{60}^{+}$ are 
in agreement with five DIBs at 9348.4, 9365.2, 9427.8,
9577.0, and 9632.1$\Angstrom$,
confirming the earlier assignment of 
the 9577 and 9632$\Angstrom$ DIBs 
to C$_{60}^{+}$ (Foing \& Ehrenfreund 1994) 
based on the absorption spectrum
recorded in a neon matrix (Fulara et al.\ 1993). 
%

There has been long-lasting interest in relating 
the DIBs to the interstellar extinction curve
-- the wavelength ($\lambda$) dependence of 
the extinction ($A_\lambda$).
An extensive exploration and understanding 
of the possible links between the DIBs 
and the interstellar extinction curve,
particularly the 2175$\Angstrom$ extinction bump and 
the far-UV (FUV) rise at $\lambda^{-1}>5.9\mum^{-1}$, 
can provide useful information on 
the origin and composition of the elusive carriers 
as well as on the physical conditions prevailing in 
the regions where they are found 
(Wu 1972, Herbig 1975, Nandy \& Thompson 1975,
Dorschner et al.\ 1977, Schmidt 1978, Danks 1980, 
Wu et al.\ 1981, Nandy et al.\ 1982,
Witt et al.\ 1983, Seab \& Snow 1984,
Krelowski et al.\ 1987, Benvenuti \& Porceddu 1989,
D\'esert et al.\ 1995, Magier et al.\ 2001, 2005,
Xiang, Li, \& Zhong 2011, Clayton 2014).

\begin{figure}
\centerline
{  
\includegraphics[width=16cm,angle=0]{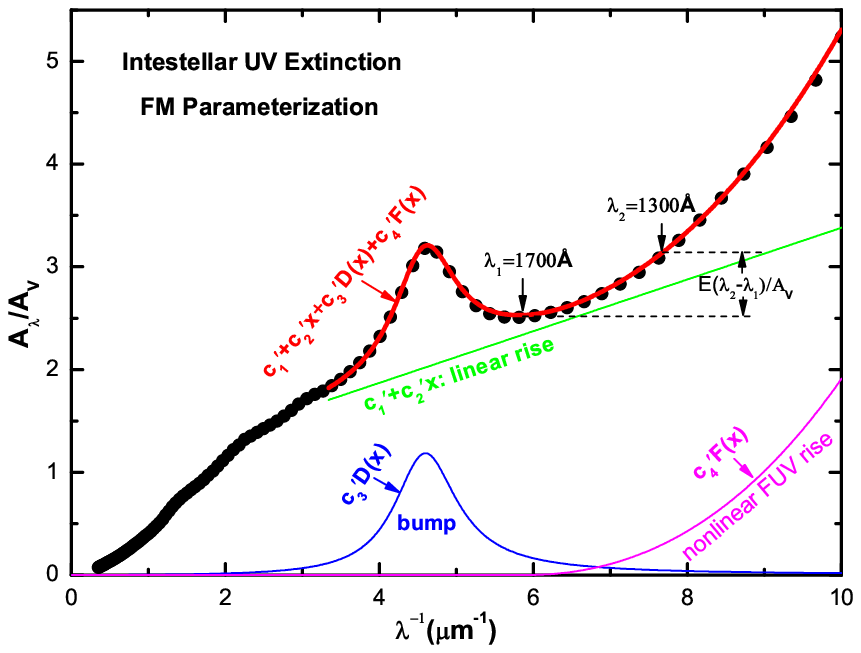}
}
\caption{\footnotesize
         \label{fig:extcurv}  
           Interstellar extinction curve 
           expressed as $A_\lambda/A_V$.
           The black circles are computed from 
           the CCM parametrization (Cardelli et al.\ 1989).
           The red line is the FM parametrization 
           at $\lambda^{-1} > 3.3\mum^{-1}$
           (Fitzpatrick \& Massa 1990; see eq.\,\ref{eq:A2AV})
           which is the sum of a linear ``background'' 
           extinction (green line),
           a Drude bump of width $\gamma$ 
           peaking at $\lambda=2175\Angstrom$
           or $\lambda^{-1}\approx4.6\mum^{-1}$ (blue line),
           and a nonlinear FUV rise (purple line)
           at $\lambda^{-1} > 5.9\mum^{-1}$.
           In the literature, the UV extinction
           is often ``measured'' by $c_2^{\prime}$,
           $c_4^{\prime}$, or $E(1300-1700)$.
           The color excess
           $E(1300-1700)\equiv A(1300\Angstrom)-A(1700\Angstrom)$ 
           is the difference between the extinction 
           at $\lambda=1300\Angstrom$ 
           and that at $\lambda=1700\Angstrom$.
           }
\end{figure}

The interstellar extinction curve
is usually expressed as $A_\lambda/\AV$
or $E(\lambda-V)/E(B-V)$,
where $\AV$ is the extinction in the visual ($V$) band,
$A_B$ is the extinction in the blue ($B$) band,
$E(\lambda-V)\equiv A_\lambda - \AV$ is the color excess
or reddening between $\lambda$ and $V$,
and $E(B-V)$ is related to $A_V$ through $R_V$,
the total-to-selective extinction ratio,
i.e., $R_V\equiv A_V/E(B-V)$.
The Galactic diffuse interstellar medium (ISM)
has an average value of $R_V\approx3.1$.
%
%
As shown in Figure~\ref{fig:extcurv}, 
the Galactic extinction curve plotted
against the inverse wavelength $\lambda^{-1}$
rises almost linearly from the near-IR to the near-UV,
with a broad absorption bump at about
$\lambda^{-1}$$\approx$4.6$\mum^{-1}$
($\lambda$$\approx$2175$\Angstrom$)
and followed by a steep rise into the FUV
at $\lambda^{-1}$$\approx$10$\mum^{-1}$,
the shortest wavelength at which the dust 
extinction has been measured.
%
%
Depending on the environment
along the line of sight,
the shape of the extinction curve,
particularly, 
the strength of the 2175$\Angstrom$ bump
and the steepness of the FUV rise, 
vary with $R_V$. 
Low-density regions usually
have a rather low value of $\RV<3.1$,
and their extinction curves exhibit 
a strong 2175$\Angstrom$ bump
and a steep FUV rise
at $\lambda^{-1}$\,$>$\,5.9$\mum^{-1}$.
Lines of sight penetrating into dense clouds,
such as the Ophiuchus or Taurus molecular clouds,
usually have $4 < \RV < 6$,
showing a weak 2175$\Angstrom$ bump,
and a relatively flat FUV rise
(see Li et al.\ 2015).




One would expect some kind of relation 
between the DIBs and the UV extinction.
First, if the DIBs and the 2175$\Angstrom$ bump
or the FUV extinction share a common carrier,
one would expect a positive correlation between them.
Second, if the carriers of the DIBs and that of
the 2175$\Angstrom$ bump and/or that of
the FUV extinction 
are produced from the same parental dust
through the same process 
(e.g., collisional fragmentation),
or the former simply originates from
the collisional grinding of the latter,
one would expect them to be related.
Finally, the DIBs could be related to
the UV extinction in an indirect way 
through the shielding of the UV starlight.
The DIB carriers can be destroyed by 
UV stellar photons. 
The UV extinction reduces the flux of 
the UV radiation and therefore shields 
the DIB carriers from being destroyed.   
In this case, the DIBs and the FUV
extinction should be positively correlated.
On the other hand, it is also possible that
the creation of the DIB carriers may require
UV photons, e.g., the carriers are possibly
ions and therefore one needs UV photons to
photoionize the carriers (e.g., see Witt 2014). 
In this case, the DIBs and the UV extinction 
should be anti-correlated.
The UV starlight is important 
in setting the balance between 
the formation and destruction 
of the DIB carriers
(Le Page et al.\ 2003, 
Ruiterkamp et al.\ 2005, 
Cox \& Spaans 2006). 


Investigating the possible correlations 
between the DIBs and the UV extinction 
could potentially provide insights into
the nature and origin of the DIB carriers
and how they respond to the local physical
environments.
%
%
About two decades ago, 
D\'esert, Jenniskens, \& Dennefeld (1995)
published a paper entitled
``Diffuse Interstellar Bands and UV Extinction Curves
-- The Missing Link''. 
In recent years, observational, experimental and 
theoretical advances in the studies of interstellar
extinction and DIBs have rapidly lead to renewed 
interest in this topic (see Cox 2011, Clayton 2014). 
%
There has been considerable progress in 
the measurements of the interstellar extinction
(e.g., see Valencic et al.\ 2004)
and DIBs (see Cami \& Cox 2014). 
There is also an improved understanding 
of the interstellar extinction models
(Weingartner \& Draine 2001, Draine \& Li 2007, 
Jones et al.\ 2013, Wang, Li \& Jiang 2015a,b)  
and of the possible carrier candidates of DIBs
(e.g., see Jones 2014, Omont 2016).
Making use of these advances, we aim at
an extensive revisit of 
the possible ``missing link'' between 
the UV extinction and DIBs. 
In an earlier paper, we have examined the relation
between the 2175$\Angstrom$ bump and DIBs
(see Xiang, Li, \& Zhong 2011) and found no correlation.
In this paper we shall focus on the relation between
the FUV extinction and DIBs.

In \S\ref{sec:history} we briefly summarize 
the previous studies on 
the possible relation between the DIBs 
and the FUV extinction.
We propose in \S\ref{sec:drude} 
the most reasonable ways to characterize 
the strength of the FUV extinction.
In \S\ref{sec:correlation} we describe 
how the sample is selected and
how the DIB strength is determined.
Also in \S\ref{sec:correlation}
we analyze the possible correlation between
the DIBs and the FUV extinction.
The results are discussed in \S\ref{sec:discussion} 
and summarized in \S\ref{sec:summary}.

\section{DIBs vs. the FUV Extinction: 
         Where Do We Stand?\label{sec:history}}
There have been a number of previous studies  
on the relation between the FUV extinction and DIBs.
Most studies compared the equivalent width (EW) or
the central depth ($A_c$) of a DIB 
with the FUV color excesses or the shape parameters 
of the extinction curve derived from 
the Fitzpatrick \& Massa (1990; hereafter FM)
decomposition scheme.

Wu (1972) compared the strengths of three DIBs
at 4430, 5780, and 5797$\Angstrom$ derived 
from the high resolution ($\simali$0.17$\Angstrom$) 
echelle spectra of 66 lines of sight
with the UV color 
$E(1200-V)\equiv A(1200) - A_V$,
where $A(1200)$ is the extinction 
at $\lambda=1200\Angstrom$ measured by 
the {\it Orbiting Astronomical Observatory} 2 (OAO-2). 
Wu (1972) found that the DIBs and the FUV extinction 
rise $E(1200-V)$ were not correlated.

Schmidt (1978) examined the strength of 
the 5780$\Angstrom$ DIB and $A(1550)$,
the FUV extinction at 1550$\Angstrom$
for 23 lines of sight. It was found that they 
were correlated, although with considerable 
intrinsic scatter. 

Recognizing that both the DIB strength 
and the UV extinction may correlate with
$E(B-V)$,\footnote{%
  In the ISM, dust and gas are well mixed
  as indicated by the relatively constant
  gas-to-extinction ratio 
  ($A_V/N_{\rm H}\approx 5.3\times 10^{-22}\magni\cm^2$;
  see Bohlin, Savage \& Drake 1978).
  Any two interstellar quantities that depend on 
  either the amount of dust or the amount of gas 
  in the line of sight will tend to correlate
  with each other to some extent.
  To explore the correlation between the DIB strength
  and the extinction parameters,  
  the common dependence on the reddening
  $E(B-V)$ has to be cancelled out.
  }
Witt, Bohlin \& Stecher (1983) argued that,
to cancel out the common correlation with
$E(B-V)$, the DIB strength and the UV extinction 
should be normalized by $E(B-V)$.
They compared the 4430$\Angstrom$ DIB
and the FUV extinction rise $E(1250-V)$
of 20 lines of sight, both divided by $E(B-V)$.
With a correlation coefficient of $r\approx -0.45$,
they found a marginally significant negative correlation.

Seab \& Snow (1984) investigated the correlation
between the DIBs at 4430, 5780, and 6284$\Angstrom$
and the FUV extinction at 1250$\Angstrom$ 
and the extinction slope between 1250$\Angstrom$ 
and 1430$\Angstrom$ of 50 lines of sight. 
No correlation was found when all quantities 
were divided by $E(B-V)$.
 
Krelowski et al.\ (1987) compared the excess 
absorption of the 4430$\Angstrom$ DIB to 
the FUV color excess $E(1200-1800)$,
the difference of the extinction at
$\lambda=1200\Angstrom$ 
and $\lambda=1800\Angstrom$. 
An anti-correlation was found 
for a combination of lines of sight 
to the Perseus OB1 and Cepheus OB3 associations.
Krelowski et al.\ (1992) found that the intensity
ratio of the 5780$\Angstrom$ DIB 
to the 5797$\Angstrom$ DIB observed 
along lines of sight toward several isolated 
interstellar clouds varies with the shape
of the UV extinction (i.e., the 2175$\Angstrom$
bump and the FUV rise).

Instead of correlating the DIB strengths with
the FUV extinction color excesses, 
D\'esert et al.\ (1995) compared the EWs of 
eight DIBs at 5707, 5780, 5797, 5850, 6177, 
6196, 6269, 6284$\Angstrom$ 
of 28 lines of sight
with the different components
of the FUV extinction
which are decomposed 
according to the FM scheme. 
In this scheme, the UV extinction
in the wavelength range of
$3.3\mum^{-1} < \lambda^{-1} < 8\mum^{-1}$
is decomposed into three distinct components:
a linear background, a Drude bump, 
and a nonlinear FUV curvature component.
This decomposition is given by the following equation:
\begin{equation}\label{eq:E2EBV}
E(\lambda-V)/E(B-V) =
c_1 + c_2\,x + c_3\,D(x;\gamma,x_0) + c_4\,F(x)~~,
\end{equation}
where $x\equiv \lambda^{-1}$ 
is the inverse wavelength ($\mu$m$^{-1}$),
the parameters $c_1$ and $c_2$ respectively 
determine the intercept and slope of
the linear ``background'' extinction,
$c_3$ determines the strength of the 2175$\Angstrom$
extinction bump which is represented by 
the Lorentz oscillator ``Drude profile'' 
$D(x;\gamma,x_0)$\,$\equiv$\,$x^2/\left[\left(x^2-x^2_0\right)^2
+ x^2\gamma^2\right]$,
$\gamma$ and $x_0$ are respectively
the FWHM and peak position of the Drude profile,\footnote{%
  In the Galactic ISM, the strength and width of
  the 2175$\Angstrom$ extinction bump vary with
  environment while its peak position is quite invariant
  (see Li 2005). 
  }
and $c_4$ determines the FUV nonlinear rise 
which is expressed as a polynomial
\begin{equation}\label{eq:FUV}
F(x) =
\left\{\begin{array}{ll}
0.5392\,\left(x-5.9\right)^2 + 0.05644\,\left(x-5.9\right)^3~,
& x\ge 5.9\mum^{-1}~~,\\
0, & x\,<\,5.9\mum^{-1}~~.\\
\end{array} \right .
\end{equation}
%
With a correlation coefficient of
$r\approx -0.44$,
D\'esert et al.\ (1995) found 
a trend for the DIB strength
[normalized by $E(B-V)$] 
to anti-correlate with 
the FUV nonlinear rise ($c_4$).
With $r\approx 0.12$,
no correlation was seen between 
the normalised DIB strength
and the linear rise in the UV ($c_2$).
They also found a trend for the normalized
DIB strength to correlate with the bump height 
$c_3/\gamma^2$ ($r\approx 0.74$)
and anti-correlate with the bump width 
$\gamma$ ($r\approx -0.40$).

Noteworthy, when expressed as $A_\lambda/A_V$,
the FM parametrization for the extinction curve
at $x>3.3\mum^{-1}$ becomes
\begin{equation}\label{eq:A2AV}
A_\lambda/A_V = c_1^{\prime} + c_2^{\prime}\,x 
              + c_3^{\prime}\,D(x;\gamma, x_0) 
              + c_4^{\prime}\,F(x) ~~~,
\end{equation}
where the parameters $c_j^{\prime}$ ($j=1, 2, 3, 4$)
are related to the FM parameters $c_j$ 
(see eq.\,\ref{eq:E2EBV}) through
\begin{equation}\label{eq:c_j}
c_j^{\prime} = \left\{\begin{array}{lr} 
c_j/R_V + 1 ~, & j=1 ~~~,\\
c_j/R_V ~, & j=2, 3, 4 ~~~.\\
\end{array}\right.
\end{equation}

Megier et al.\ (2001) studied the relation between 
the two major DIBs, 5780 and 5797$\Angstrom$, 
of 70 lines of sight, and different color excesses
$E(\lambda_1-\lambda_2)$ in the wavelength
range of $1260\Angstrom<\lambda<3200\Angstrom$,
with both the DIB strengths and the UV color excesses
normalized by $E(B-V)$.
With a step of 20$\Angstrom$ adopted for
$\lambda_1$ and $\lambda_2$, 
there were 4753 different possible
($\lambda_1$, $\lambda_2$) pairs,
and thus, 4753 different color excesses.
Megier et al.\ (2001) found that 
the correlation patterns of 
the two DIBs were different:
showing an overall better correlation 
than the 5780$\Angstrom$ DIB,
the 5797$\Angstrom$ DIB
correlates best with the color excesses
when at least one of the wavelengths 
is in the FUV region, while the 5780$\Angstrom$ DIB 
exhibits good correlation with the color excesses 
when one of the wavelengths falls close to 
the 2175$\Angstrom$ bump.
The ratio of these two DIBs is best correlated with 
the FUV extinction rise.
Megier et al.\ (2005) further extended 
the same analysis to 11 DIBs of 49 lines of sight.
They found that, while most of the DIBs correlate 
positively with the extinction in the neighbourhood 
of the 2175$\Angstrom$ bump,
the correlation with colour excesses in other parts 
of the extinction curve is more variable from one DIB
to another: some DIBs (5797, 5850 and 6376$\Angstrom$) 
correlate positively with the overall slope of the extinction curve, 
while others (5780 and 6284$\Angstrom$) exhibit negative correlation. 

We conclude that,
although whether (and how) the DIB strengths are 
related to the FUV extinction has been a topic of 
extensive research for over four decades,
no consensus has been reached and
contradicting conclusions have been drawn in the literature.
This seems to be largely caused by the inconsistencies 
in characterizing the strength of the FUV extinction.

\section{How to Characterize the FUV Extinction?
         \label{sec:drude}}
As summarized in \S\ref{sec:history},
whether the carriers of DIBs are related to 
the FUV extinction has been a topic of 
extensive studies for over four decades.
Contradicting conclusions have been drawn 
in the literature, largely caused 
by the inconsistencies in characterizing 
the FUV extinction. 
In the majority of previous work on this subject, 
the FUV extinction was characterized 
by a few chosen quantities. 
These were either selected {\it extinction}
[Schmit (1978): $A(1550)$; 
Seab \& Snow (1984): $A(1250)$],
{\it color excesses}
[Wu (1972): $E(1200-V)$;
Witt et al.\ (1983): $E(1250-V)$;
Krelowski et al.\ (1987): $E(1200-1800)$;
Megier et al.\ (2001, 2005): 
$E(\lambda_1-\lambda_2)$ where 
$1260\Angstrom\simlt\lambda_1 < \lambda_2\simlt3200\Angstrom$],
or {\it parameters} 
(e.g., $c_2$, $c_4$; see eq.\,\ref{eq:E2EBV})
derived by fitting the extinction curve 
with the FM parametrization 
(D\'esert et al.\ 1995).
Correlation between these quantities 
and the DIBs was then sought. 
In view of the various standards 
in characterizing the strength of 
the FUV extinction, an important question 
one need to address is:
what is the most reasonable measure of 
the FUV extinction?
We note that, while the FM parametrization 
provides an excellent mathematical description
of the UV extinction at $\lambda^{-1}>3.3\mum^{-1}$,
the distinction between the linear rise 
(measured by $c_1$ and $c_2$)
and the FUV non-linear rise (measured by $c_4$)
probably has little physical significance 
since there is no substance known that shows 
the corresponding extinction of any of them.

\begin{figure}
\centerline
{
\includegraphics[width=16cm,angle=0]{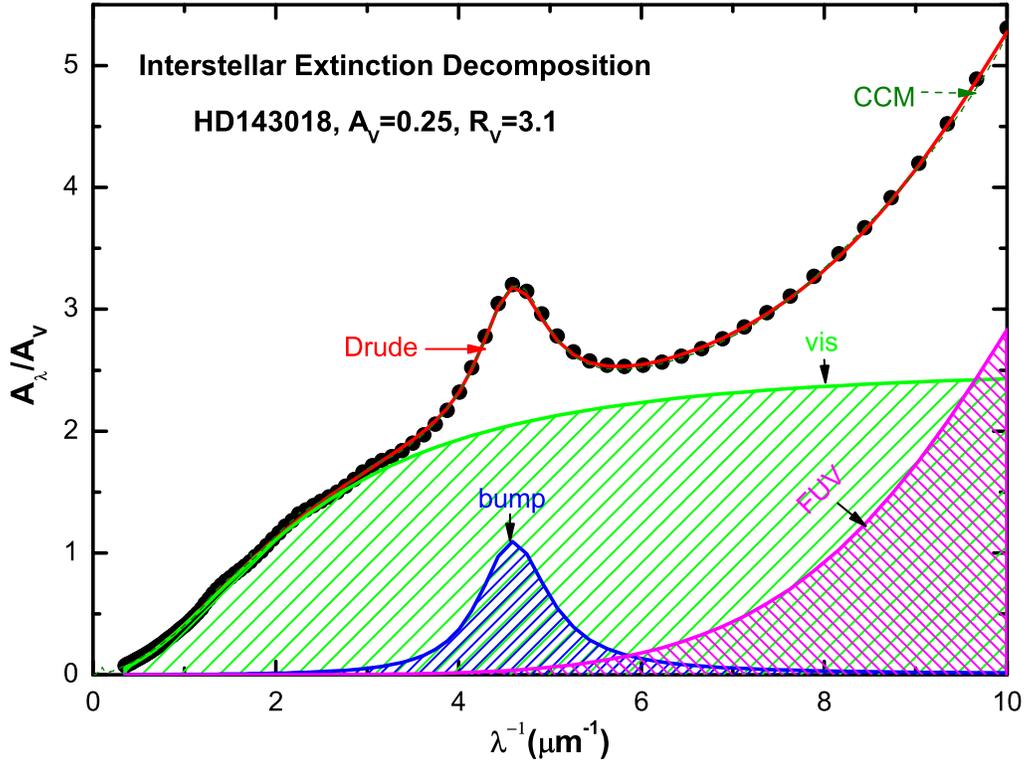} 
}
\caption{\footnotesize
           \label{fig:drude}
           Decomposition of the extinction curve
           (black circles) observed toward HD\,143018 
           into three components: 
           the visual component (green line),
           the bump component (blue line),
           and the FUV component (purple line).   
           Each component is represented by
           a Drude function.
           The solid red line labelled with ``Drude''
           is the sum of 
           all these three Drude components.
           The dashed green line is the CCM fit
           with $R_V=3.1$. 
           We propose to take the shaded area 
           under each extinction component as
           a measure of the quantity of its carrier
           (see \S\ref{sec:drude}).
           }
\end{figure}

As illustrated in Figure~\ref{fig:drude},
we propose to decompose the interstellar extinction 
curve into three Drude-like functions: 
\beqa \label{eq:drude} 
\Alambda/\AV & = &
\frac{a_3}{\left(\lambda/\lambdaVIS\right)^{a_1}
+ \left(\lambdaVIS/\lambda\right)^{a_1} + a_2}
\nonumber
\\ 
& + & \frac{a_5}{\left(\lambda/0.2175\right)^{2}
+ \left(0.2175/\lambda\right)^{2} + a_4}
\nonumber
\\
& + & \frac{a_8}{\left(\lambda/\lambdaFUV\right)^{a_6}
+ \left(\lambdaFUV/\lambda\right)^{a_6} + a_7} ~~~, 
\eeqa
where $\lambda$ is in $\mu$m,
and the parameters $a_1$, $a_2$, ..., $a_8$
are dimensionless. 
In eq.\,\ref{eq:drude}, the first term 
on the right-hand side represents 
the near-IR/visible extinction,
the second term represents 
the 2175$\Angstrom$ extinction bump,
and the third term accounts for 
the FUV extinction.
The parameters $a_1$, $a_2$, ..., and $a_8$
are not all independent. 
At $\lambda=0.55\mum$, the left-hand side of
eq.\,\ref{eq:drude} becomes unity and therefore
one derives $a_8$ from $a_1$, $a_2$, ..., and $a_7$:
\beq
\label{eq:a8}
a_8 = \left\{
1 - \frac{a_3}{\left(0.55/\lambdaVIS\right)^{a_1}
+ \left(\lambdaVIS/0.55\right)^{a_1} + a_2}
- \frac{a_5}{6.551+a_4} 
\right\}
\times
\left\{
\left(0.55/\lambdaFUV\right)^{a_6}
+ \left(\lambdaFUV/0.55\right)^{a_6} + a_7
\right\} ~.
\eeq

\begin{figure}
\center
{
\includegraphics[width=9cm,angle=0]{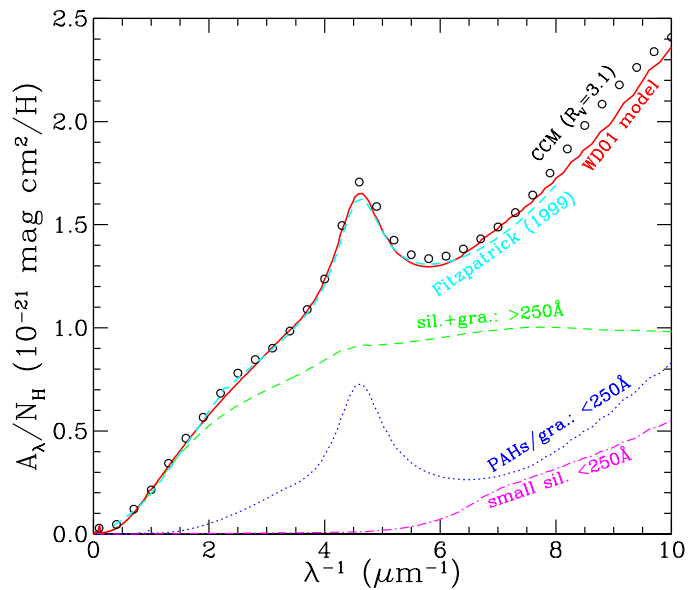} 
}
\caption{\footnotesize
         \label{fig:wd01decomp}
         Comparison of the extinction curve of
         the WD01 silicate-graphite-PAH model 
         (solid red line) with the Galactic average 
         extinction represented by the CCM parametrization 
         of $R_V=3.1$ (black open circles)
         and the Fitzpatrick (1999) formula 
         (cyan dashed line).
         The WD01 model extinction is separated
         into three components: the extintion caused
         by ``classical'' silicate and graphite grains
         of radii $a>250\Angstrom$ (green dashed line),
         the extinction produced by PAHs or ultrasmall
         graphitic grains of radii $a<250\Angstrom$
         (blue dotted line),
         and the extinction arising from  ultrasmall
         silicate grains of radii $a<250\Angstrom$.   
         }
\end{figure}

Originally introduced by Pei (1992),
the above-described ``Drude'' decomposition scheme
has been successfully applied to derive 
the extinction curves of the host galaxies
of $\gamma$-ray bursts (see Li et al.\ 2008, 
Liang \& Li 2009, 2010).
The physical basis of this decomposition scheme
lies in the interstellar grain size distribution.
As early as 1973, Greenberg (1973) suggested to
separate the interstellar extinction curve into
three parts: the near-IR/visual part dominated 
by the sub-$\mu$m-sized ``classical'' grains
of $a\simgt0.05\mum$, the 2175$\Angstrom$ bump 
caused by grains much smaller than 
the so-called ``classical'' sizes,
and the FUV rise produced by small grains with
radii $a<0.01\mum$. 
This can be quantitatively verified in terms of
the Weingartner \& Draine (2001; hereafter WD01) model 
which consists of a mixture of amorphous silicate 
and carbonaceous grains ranging from 
a few angstroms to a few tenth micrometers. 
The carbonaceous grain population extends from grains 
with graphitic properties at radii $a\simgt0.01\mum$, 
down to particles with PAH-like properties 
at very small sizes (Li \& Draine 2001a).
As demonstrated in Figure~\ref{fig:wd01decomp},
the extinction arising from ``classical'' silicate
and graphite grains of $a\simgt250\Angstrom$\footnote{%
   Li \& Draine (2001a) classified interstellar grains
   into two categories: ``classical'' grains with
   radii $a\simgt250\Angstrom$ and ultrasmall grains
   with $a\simlt250\Angstrom$. 
   The size $a=250\Angstrom$ was chosen 
   since when exposed to the general interstellar 
   radiation field (Mathis et al.\ 1983), 
   ultrasmall grains with $a\simlt250\Angstrom$
   undergo transient heating by individual starlight
   photons while ``classical'' grains with $a\simgt250\Angstrom$   
   attain an equilibrium tmperature.
   }
indeed dominates the near-IR/visual part of 
the observed extinction curve 
and saturates at $\lambda>4\mum^{-1}$.
The reason for this saturation is that the change
in curvature of the calculated extinction curve 
from concave to convex at $\simali$2.3$\mum^{-1}$
is a dominant signature of grains with an average 
grain radii of $\simali$0.05--0.2$\mum$ (Greenberg 1968).
Also prominent in Figure~\ref{fig:wd01decomp}
is that the 2175$\Angstrom$ bump is produced
by PAHs and ultrasmall graphitic grains of 
$a\simlt250\Angstrom$,
while ultrasmall silicate grains of 
$a\simlt250\Angstrom$ mainly absorb at the FUV.
The sum of these three parts closely
reproduces the observed extinction curve
represented by the CCM formula of $R_V=3.1$
and Fitzpatrick (1999).
A comparison of the WD01 model separation
(see Figure~\ref{fig:wd01decomp}) 
with the Drude decomposition 
(see Figure~\ref{fig:drude}) 
reasonably justifies the decomposition scheme:
the near-IR/visual extinction results from
``classical'', sub-$\mu$m-sized grains
while the FUV extinction rise is caused
by a population of grains of different sizes
(i.e., ultrasmall grains with $a\simlt250\Angstrom$).
The major difference between the Drude decomposition 
and the WD01 model separation is that, 
while the former contains a separate bump component 
which contributes little to the FUV, 
the latter has the FUV extinction rise 
appreciably contributed by the bump carrier
(i.e., PAHs and ultrasmall graphitic grains).
Finally, we note that the justification of
the Drude decomposition scheme would not only
be provided by the WD01 silicate-graphite-PAH model,
but by other models as well 
(e.g., see D\'esert et al.\ 1990, 
Li \& Greenberg 1997, 
Schnaiter et al.\ 1998,
Jones et al.\ 2013).

We further propose to use $\AFUVint$,
the area of the FUV component 
integrated over $\lambda^{-1}$, 
as a measure of the FUV extinction
\beq
\label{eq:AFUVint}
\AFUVint/\AV \equiv \int_{0.1\,\mu {\rm m}}^{\infty}
\frac{a_8}{\left(\lambda/\lambdaFUV\right)^{a_6}
+ \left(\lambdaFUV/\lambda\right)^{a_6} + a_7}\,d\lambda ~~.
\eeq
The integration is made over 
$0<\lambda^{-1}<10\mum^{-1}$, 
with the upper limit of the integral
being intermediate between the short-wavelength limit 
($\simali$0.12$\mum$) of 
the {\it International Ultraviolet Explorer} (IUE) satellite
and the Lyman limit at 0.0912$\mum$.
It is easy to show that $\AFUVint$ measures 
the column density of the FUV absorber 
(see Purcell 1969):
\beq
\label{eq:NFUV}
\AFUVint = \int_{0.1\,\mu {\rm m}}^{\infty}
\AFUV(\lambda)\,d\lambda \propto \NFUV ~~,
\eeq
where $\NFUV$ is the column density of the FUV absorber. 
Noteworthy, the column density of the carrier 
of a given DIB can be determined from its equivalent width 
$\WDIB$ through
\beq
\label{eq:NDIB}
\NDIB \approx 1.13\times 10^{20}\cm^{-2} 
\left(\WDIB/\Angstrom\right)
\left(\lambda/\Angstrom\right)^{-2}\,\fdib^{-1} ~~,
\eeq
where $\fdib$ is the oscillator strength of the transition 
associated with the DIB (see Herbig 1993).
Therefore, one can compare $\AFUVint$ with $\WDIB$
as both quantities linearly determine 
the column densities of their respective carriers.
%
We note that, strictly speaking, 
the perfect linear proportionality 
between $\AFUVint$
and $\NFUV$ (see eq.\,\ref{eq:NFUV})
or between $\WDIB$ and $\NDIB$
(see eq.\,\ref{eq:NDIB}) is strictly valid 
only if the intrinsic physical and chemical 
properties of the carriers 
of the FUV extinction or DIBs
do not vary much in different 
interstellar environments.
This is because $\WDIB\propto\fdib\times\NDIB$
and $\AFUVint\propto F\times\NFUV$,
where the dimensionless factor $F$
depends on the shape of the FUV extinction carrier
and the static (zero-frequency) dielectric constant 
$\varepsilon_0$ of the carrier material  (Purcell 1969).
If the intrinsic chemical and structural properties 
of the carriers of the FUV extinction or DIBs
differ substantially in different environments, 
$\varepsilon_0$ and $\fdib$ may not be invariant. 




\section{Correlations between DIBs and 
         the FUV Extinction}\label{sec:correlation}
To examine whether (and how) DIBs correlate 
with the FUV extinction, we compile an as large as possible
set of sightlines for which both the interstellar 
extinction curves and DIBs have been measured.
Also, in order for our analysis of the correlation
between DIBs and the FUV extinction to be statistically
significant, we require that each DIB has been measured 
in at least five sightlines. 
To this end, we have collected the following 40 DIBs
for 97 sightlines: 
$\lambda$4428/4430$\Angstrom$, 
$\lambda$4501$\Angstrom$, 
$\lambda$4726$\Angstrom$, 
$\lambda$4762/4763$\Angstrom$,
$\lambda$5487$\Angstrom$, 
$\lambda$5544$\Angstrom$, 
$\lambda$5705$\Angstrom$, 
$\lambda$5707$\Angstrom$, 
$\lambda$5763$\Angstrom$, 
$\lambda$5766$\Angstrom$, 
$\lambda$5773$\Angstrom$, 
$\lambda$5776$\Angstrom$,
$\lambda$5778$\Angstrom$,  
$\lambda$5780$\Angstrom$, 
$\lambda$5793$\Angstrom$, 
$\lambda$5795$\Angstrom$, 
$\lambda$5797$\Angstrom$, 
$\lambda$5809$\Angstrom$, 
$\lambda$5819$\Angstrom$, 
$\lambda$5829$\Angstrom$, 
$\lambda$5844$\Angstrom$, 
$\lambda$5849/5850$\Angstrom$,
$\lambda$6010$\Angstrom$, 
$\lambda$6065$\Angstrom$, 
$\lambda$6090$\Angstrom$, 
$\lambda$6113$\Angstrom$, 
$\lambda$6195/6196$\Angstrom$,
$\lambda$6203$\Angstrom$, 
$\lambda$6204.5$\Angstrom$, 
$\lambda$6234$\Angstrom$, 
$\lambda$6269/6270$\Angstrom$,
$\lambda$6284$\Angstrom$, 
$\lambda$6376$\Angstrom$,
$\lambda$6379$\Angstrom$,
$\lambda$6425$\Angstrom$, 
$\lambda$6439$\Angstrom$, 
$\lambda$6521$\Angstrom$, 
$\lambda$6613/6614$\Angstrom$,
$\lambda$6660/6661$\Angstrom$, and 
$\lambda$6699$\Angstrom$.
The $\lambda$4428$\Angstrom$ DIB,
a DIB so far detected at the shortest wavelength,  
is sometimes known as the $\lambda$4430$\Angstrom$ DIB.
In this work we consider them as the same DIB
and label them ``$\lambda$4428/4430$\Angstrom$''.
%
Similarly, the DIBs at $\lambda$4762$\Angstrom$,
$\lambda$5849$\Angstrom$,
$\lambda$6195$\Angstrom$,
$\lambda$6269$\Angstrom$, 
$\lambda$6613$\Angstrom$, and 
$\lambda$6660$\Angstrom$
are often also respectively referred to as 
the $\lambda$4763$\Angstrom$,
$\lambda$5850$\Angstrom$,
$\lambda$6196$\Angstrom$,
$\lambda$6270$\Angstrom$, 
$\lambda$6614$\Angstrom$, and
$\lambda$6661$\Angstrom$ DIBs.
In this work we also treat them
as single DIBs and label them
``$\lambda$4762/4763$\Angstrom$'',
``$\lambda$5849/5850$\Angstrom$'',
``$\lambda$6195/6196$\Angstrom$'', 
``$\lambda$6269/6270$\Angstrom$'', 
``$\lambda$6613/6614$\Angstrom$'', and
``$\lambda$6660/6661$\Angstrom$''.
In addition, the two DIBs at $\lambda$6376$\Angstrom$ 
and $\lambda$6379$\Angstrom$ may blend.
To examine this effect, we have also 
considered grouping these two DIBs 
into one DIB which is referred to as 
``$\lambda$6376$\Angstrom$/6379$\Angstrom$".
We derive similar correlation results for  
the ``blended'' $\lambda$6376$\Angstrom$/6379$\Angstrom$ DIB
as that if the $\lambda$6376$\Angstrom$ DIB
and the $\lambda$6379$\Angstrom$ DIB are
treated separately.
%

\begin{figure}
\center
{
\includegraphics[width=16cm,angle=0]{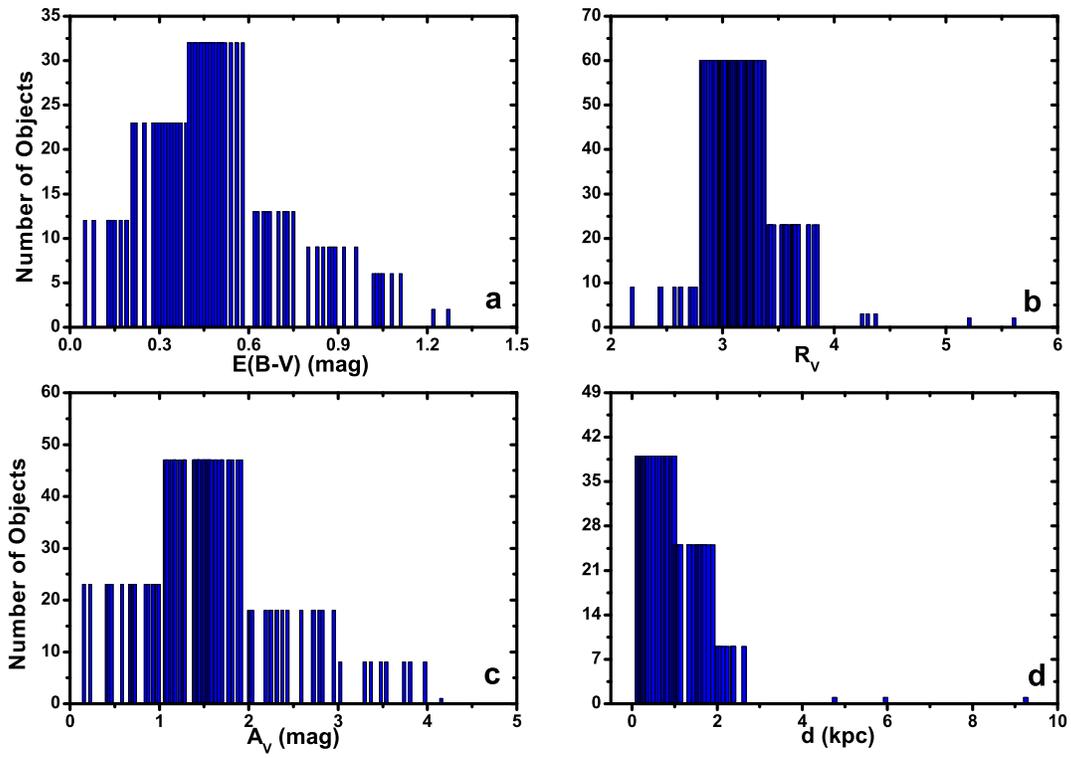} 
}
\caption{\footnotesize
         \label{fig:histogram}
         Histogram of (a) the reddening $E(B-V)$, 
         (b) the total-to-selective extinction ratio $R_V$,
         (c) the visual extinction $A_V$, and
         (d) the distance $d$ from Earth to the cloud
         for 97 lines of sight. 
         }
\end{figure}

We stress that, apparently,
the selected 40 DIBs have been 
detected in many other sources as well
(in addition to the selected 97 lines of sight).
Also, many more DIBs (in addition to 
the selected 40 DIBs) have also been seen 
in the selected 97 lines of sight.
However, by satisfying 
the requirement that (1) for any given DIB,
it must have been detected 
in at least five sightlines,
and (2) for any given line of sight,
the UV extinction curve must have 
been measured, we are finally left with 
a sample of 40 DIBs and 97 sightlines.
In Figure~\ref{fig:histogram} we present 
the histograms of $E(B-V)$, $R_V$, $A_V$,
and $d$, the distance from Earth 
to the cloud for the selected 97 lines of sight. 
It can be seen that the majority of these sources
has $E(B-V) < 1\magni$ (for 89/97 of the sources), 
$R_V<3.5$ (for 75/97 of the sources), 
$A_V<3\magni$ (for 88/97 of the sources)
and $d<2\kpc$ (for 64/76 of the sources),\footnote{%
    For the 97 clouds, we only know 
    the distance to 76 of them. 
    }
with a median value of 
$\langle E(B-V) \rangle\approx 0.46\magni$,
$\langle R_V \rangle\approx 3.2$, 
$\langle A_V \rangle\approx 1.5\magni$,
and $\langle d \rangle\approx 0.99\kpc$.
This demonstrates that the majority 
of the interstellar clouds considered 
in this work is diffuse or translucent
in nature, not dense.\footnote{%
   The ISM is generally classified into 
   three phases (see Snow \& McCall 2006):
   the cold neutral medium, 
   the warm ionized medium or 
   warm neutral medium, 
   and the hot ionized medium.
   The cold neutral medium itself 
   contains a variety of cloud types
   (e.g., diffuse clouds, translucent clouds,
   and molecular clouds).
   In diffuse clouds,
   hydrogen is mostly in atomic form and carbon is mostly
   in ionized form (C$^{+}$). They typically have a total
   visual extinction of $\AV$\,$\simali$0--1$\magni$, 
   and a hydrogen number density of
   $\nH$\,$\simali$10--500$\cm^{-3}$.
   In translucent clouds, 
   hydrogen is mostly in molecular (H$_2$)
   and the transition of carbon 
   from ionized (C$^{+}$) to atomic (C) 
   or molecular (CO) form takes place.
   These clouds have 
   $\AV$\,$\simali$1--5$\magni$ and
   $\nH$\,$\simali$500--5000$\cm^{-3}$.
   In molecular clouds,
   carbon becomes almost completely molecular.
   They are dense ($\nH$\,$\simgt$\,$10^4\cm^{-3}$)
   and subject to large extinction 
   ($\AV$\,$>$\,5--10$\magni$).
   %
   }
In this sense, the intrinsic chemical and structural 
properties of the carriers of the far-UV extinction 
or DIBs are not expected to vary substantially 
among these 97 interstellar clouds. 
Therefore, it is reasonable to expect 
a more or less linear relationship
between $\AFUVint$ and $\NFUV$
(see eq.\,\ref{eq:NFUV})
or between $\WDIB$ and $\NDIB$
(see eq.\,\ref{eq:NDIB}).

We construct a set of 97 ``observed''
extinction curves as follows.
We take the FM parameters 
($c_1$, $c_2$, $c_3$, $c_4$, 
$x_{\rm o}$, $\gamma$) 
derived in the literature 
for the UV extinction curves 
at $\lambda^{-1} > 3.3\mum^{-1}$
of all these 97 lines of sight.  
These parameters as well as 
$E(B-V)$ and $R_V$ are tabulated
in Table~\ref{tab:FMPara}.
%
The optical/near-IR extinction curves
at $0.3 < \lambda^{-1} < 3.3\mum^{-1}$
are computed from the $R_V$-based 
CCM parametrization.
We then smoothly join the UV extinction
at $\lambda^{-1} > 3.3\mum^{-1}$
to the optical/near-IR extinction. 
%
For each sightline, by following the approach
elaborated in \S\ref{sec:drude} (see eq.\,\ref{eq:drude}) 
we decompose the ``observed'' extinction curve into 
three Drude functions and obtain 
the integrated FUV extinction $\AFUVint$.
%
Following WD01,
we evaluate the extinction at 
100 wavelengths $\lambda_i$, 
equally spaced in $\ln\lambda$.
We use the Levenberg-Marquardt method
(Press et al.\ 1992) to minimize $\chi^2$ 
which gives the error 
in the decompositional fit:
\begin{equation}
\label{eq:chisq}
\chi^2 = \sum_i \frac{\left\{
      \left( A_{\lambda,i}/A_V\right)_{\rm obs} 
       - \left( A_{\lambda,i}/A_V\right)_{\rm mod}
      \right\}^2} {\sigma_i^2}~~~,
\end{equation}
where $\left( A_{\lambda,i}/A_V\right)_{\rm obs}$ 
is the ``observed'' extinction 
at wavelength $\lambda_i$,
$\left( A_{\lambda,i}/A_V\right)_{\rm mod}$ 
is the extinction computed
from the decompositional fit 
at wavelength $\lambda_i$
(see eq.\,\ref{eq:drude}), 
and $\sigma_i$ is the weight.  
Following WD01,
we take 
$\sigma_i =0.03\times\left( A_{\lambda,i}/A_V\right)_{\rm obs}$
for $1.1 < \lambda^{-1} < 10\mum^{-1}$ and
$\sigma_i = 0.1\times\left( A_{\lambda,i}/A_V\right)_{\rm obs}$ 
for $\lambda^{-1} < 1.1\mum^{-1}$.
%
The Drude parameters $a_1$, $a_2$, ..., $a_8$,
$\lambdaVIS$, and $\lambdaFUV$
as well as $\AFUVint$ and $\chi^2$ 
are tabulated in Table~\ref{tab:DrudePara}.
Also shown in Table~\ref{tab:DrudePara}
are $\Abumpint$, the wavelength-integrated extinction
of the 2175$\Angstrom$ bump and $\AVISint$,
the wavelength-integrated visual/near-IR extinction
which are defined as:
\beq
\label{eq:Abumpint}
\Abumpint/\AV \equiv \int_{0.1\,\mu {\rm m}}^{\infty}
\frac{a_5}{\left(\lambda/0.2175\right)^{2}
+ \left(0.2175/\lambda\right)^{2} + a_4}\,d\lambda ~~.
\eeq
\beq
\label{eq:AVISint}
\AVISint/\AV \equiv \int_{0.1\,\mu {\rm m}}^{\infty}
\frac{a_3}{\left(\lambda/\lambdaVIS\right)^{a_1}
+ \left(\lambdaVIS/\lambda\right)^{a_1} + a_2}\,d\lambda ~~.
\eeq

\begin{figure}
\centerline
{
\includegraphics[width=18cm,angle=0]{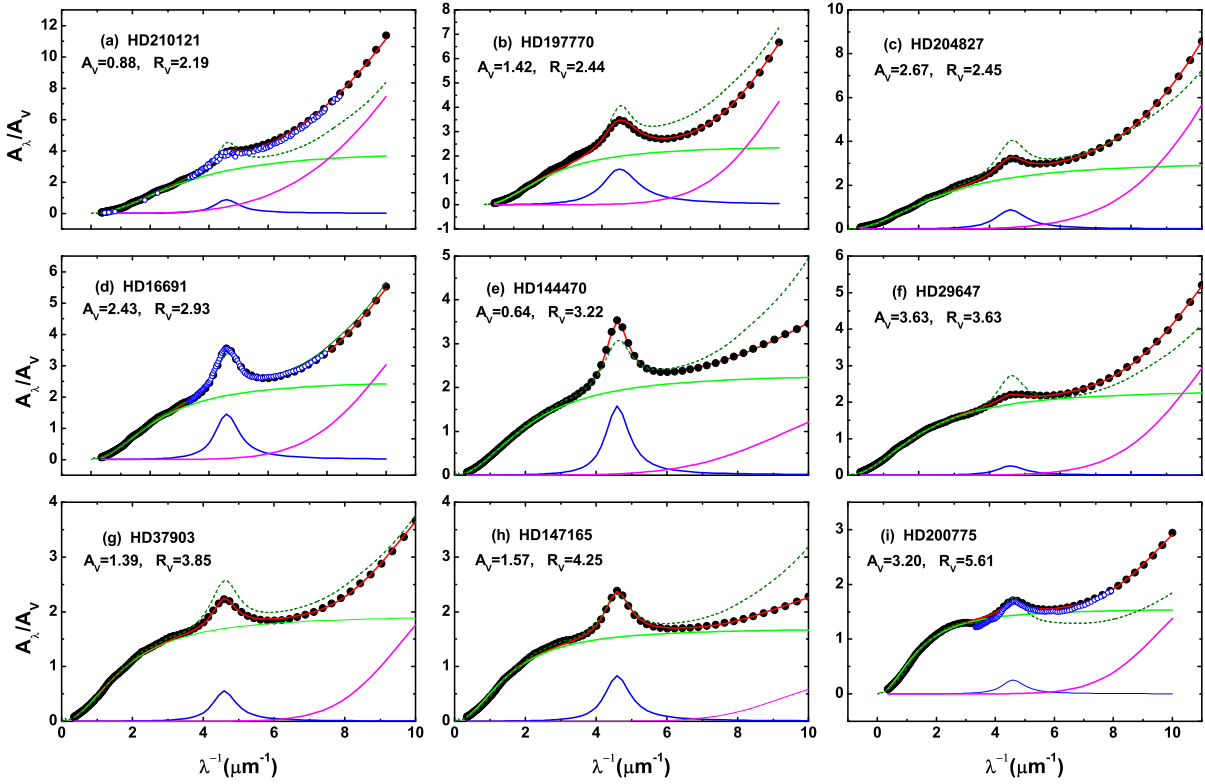} 
}
\caption{\footnotesize
         \label{fig:extcurvFit}
         Decomposing the ``observed'' extinction curves
         (black circles)
         of nine representative lines of sight 
         into three Drude-like components:
         the near-IR/visual component (solid green line),
         the 2175$\Angstrom$ bump (solid blue line),
         and the FUV rise (solid magenta line).
         The solid red line is the sum of all
         these three components.
         The dashed green line is calculated from
         the CCM parametrization with the corresponding
         $R_V$ value observationally determined for
         that specific line of sight.
         The observed extinction curves of 
         the selected nine lines of sight
         range from normal-looking CCM-type curves
         (e.g., d: HD\,16691 with $R_V=2.93$)
         to curves which substantially deviate
         from the CCM representation by showing
         a very steep FUV rise (e.g., a: HD\,210121)
         or a flat FUV rise (e.g., h: HD\,147165,
         i: HD\,200775).
         These lines of sight also have widely
         varying bump strengths, ranging from
         very weak bumps (e.g., a: HD\,210121,
         f: HD\,29647, i: HD\,200775)
         to bumps stronger than the CCM representation
         (e.g., e: HD\,144470).
         Note that the ``observed'' extinction curves
         shown here are not really the original 
         observational data, but approximated by 
         the FM parametrization 
         at $\lambda^{-1} > 3.3\mum^{-1}$
         and the CCM parametrization 
         at $\lambda^{-1} < 3.3\mum^{-1}$.
         For illustration, we also show 
         the original IUE data (blue circles) 
         for three representative lines of sight:
         HD\,210121 (a), HD\,16691 (d), 
         and HD\,200775 (i).
         }
\end{figure}

In Figure~\ref{fig:extcurvFit} we show 
the decompositional fitting results for
nine representative sightlines 
which exhibit a wide range of extinction curves. 
At $\lambda^{-1} > 3.3\mum^{-1}$,
the ``observed'' extinction curves of 
these lines of sight range from 
normal-looking CCM-type curves
(e.g., HD\,16691 for which the observed 
extinction curve closely agrees with 
the CCM parametrization of $R_V=2.93$)
to curves which substantially deviate
from the CCM formulae by showing
either a very steep FUV rise (e.g., HD\,210121)
or a flat FUV rise (e.g., HD\,147165, HD\,200775).
These lines of sight also have widely
varying bump strengths, ranging from
very weak bumps (e.g., HD\,210121,
HD\,29647, HD\,200775)
to bumps considerably stronger than 
predicted from $R_V$-based CCM formulae (e.g., HD\,144470).
While the CCM parametrization could not fit
the ``observed'' extinction curves 
at $3.3 < \lambda^{-1} <  10\mum^{-1}$
of many sightlines,
the Drude decomposition technique closely reproduces
the extinction curves of all 97 sightlines. 
%

We note that, strictly speaking, 
the ``observed'' extinction curves
shown in Figure~\ref{fig:extcurvFit}
are not really the original observational
data, but approximated by 
the FM parametrization 
at $\lambda^{-1} > 3.3\mum^{-1}$
and the CCM parametrization 
at $\lambda^{-1} < 3.3\mum^{-1}$.
To verify this parametrized approximation,
we also show the original IUE data  
for three representative lines of sight:
HD\,210121 (Fitzpatrick \& Massa 2007), 
HD\,16691 (Aiello et al.\ 1988), 
and HD\,200775 (Aiello et al.\ 1988).
While the extinction curve of HD\,16691
is well described by the CCM formula
and closely resembles the average 
Galactic extinction law, 
HD\,210121 exhibits a steep far-UV 
extinction rise and  HD\,200775 shows
a flat far-UV extinction.
As the FM parametrization closely
fits the IUE data, it is not surprising
that the Drude decompositional model
is also in close agreement with the IUE 
data (see Figure~\ref{fig:extcurvFit}a,d,i).
%

\begin{figure}
\centerline
{
\includegraphics[width=18cm,angle=0]{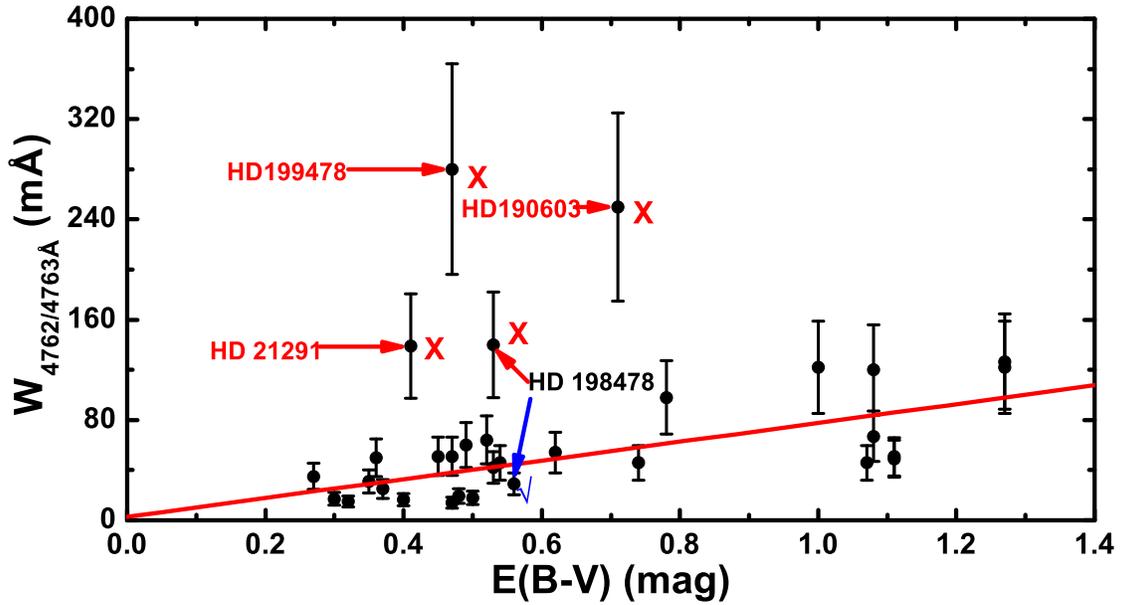} 
}
\caption{\footnotesize
             \label{fig:4762EWvsEBV}
             $W_{4762\Angstrom}$ vs. $E(B-V)$.
             The $\lambda$4762/4763$\Angstrom$ DIB
             has been observed only once 
             in HD\,21291, HD\,190603, and HD\,199478. 
             The reported EWs for these three sightlines 
             (Herbig 1975) are not adopted 
             (denoted by red ``$\times$'')
             since they deviate from the $\WDIB$--$E(B-V)$ 
             linear relation (red line) by more than $3\sigma$.
             This DIB has been observed twice in HD\,198478
              (Herbig 1975, Thorburn et al.\ 2003).
             The one measured by Thorburn et al.\ (2003)
             is adopted (denoted by blue ``$\surd$'')
             since their EW is within 3$\sigma$ 
             of that given by the $\WDIB$--$E(B-V)$ linear relation,
             while the one determined by Herbig (1975) 
             is not adopted (denoted by red ``$\times$'')
             since it deviates from 
             the $\WDIB$--$E(B-V)$ linear relation 
             by more than $3\sigma$.
             Note that, following Friedman et al.\ (2011),
             the best-fit red line has not been constrained
             to go through the origin. 
             }  
\end{figure}

\begin{figure}
\centerline
{
\includegraphics[width=18cm,angle=0]{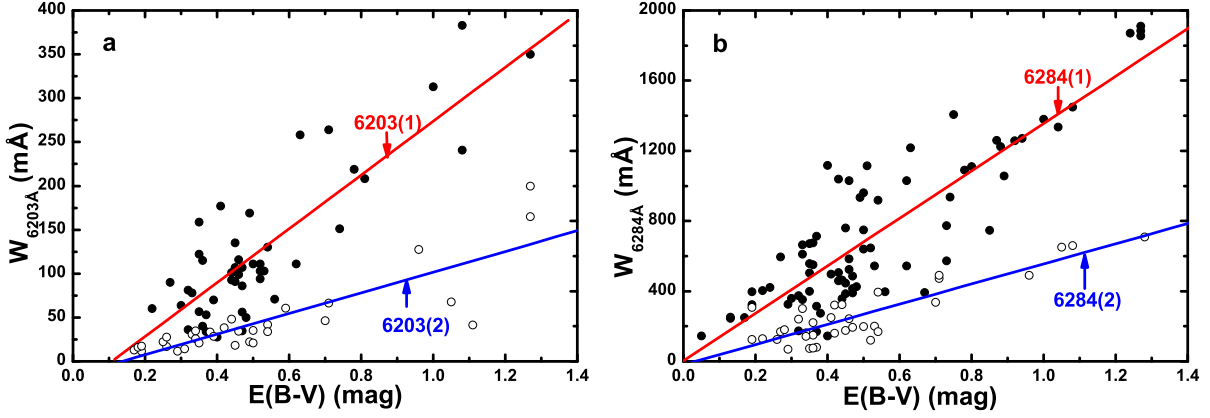} 
}
\caption{\footnotesize
             \label{fig:6203+6284EWvsEBV}
             Left panel (a): $W_{\rm 6203\,\AA}$ vs. $E(B-V)$.
             The EW data for this DIB appear to
             fall into two groups (open and filled circles) 
             each of which exhibit a different proportionality
             between $W_{6203\Angstrom}$ and $E(B-V)$.
             We label them ``6203(1)'' and ``6203(2)''.
             Following Friedman et al.\ (2011),
             the best-fit line has not been constrained
             to go through the origin. 
             Right panel (b):  Same as (a) but for
             the $\lambda$6284$\Angstrom$ DIB. 
             }
\end{figure}

The DIB EWs are also taken from the literature.
Unfortunately, there is no concensus in the DIB
community on how to derive the DIB EW. 
For the same DIB, if integrated over different
wavelength limits, one would obtain different EWs
[e.g., in determining the EW of 
the $\lambda$5797$\Angstrom$ DIB,
Friedman et al.\ (2011) performed the integration
by extending to the blue wing of the DIB, 
while for some sightlines Galazutdinov et al.\ (2004) 
did not include the blue wing; therefore, the EW of 
this DIB derived by Friedman et al.\ (2011) 
for some sightlines was 
appreciably larger than that of 
Galazutdinov et al.\ (2004)].
It is rather common that different EWs have 
been reported by different authors
for the very same DIB 
along the very same sightline 
(see Tables~\ref{tab:4428}--\ref{tab:6699}).
In some cases the reported EWs differ
substantially from one another.
It is therefore not trivial to decide 
which one is the ``true'' EW.
We employ the following criteria to 
determine the EW of a DIB of a given 
line of sight from the literature: 
\begin{itemize}
\item For each of the selected 40 DIBs, we collect 
          all the EWs ($\WDIB$) reported in the literature 
          for the sightlines of inerest here
          (i.e., they must be among the selected 97 sightlines). 
          As illustrated in Figure~\ref{fig:4762EWvsEBV},
          we plot $\WDIB$ against $E(B-V)$ 
          for the $\lambda$4762/4763$\Angstrom$ DIB 
          and derive a linear relation 
          between $\WDIB$ and $E(B-V)$. 
          A linear relation between $W_{\rm DIB}$ and $E(B-V)$       
          is expected as both quantities are proportional to
          the amount of interstellar materials 
          along the line of sight.
          It is true that the linearity 
          between $\WDIB$ and $E(B-V)$
          may break down for DIBs in molecular clouds. 
          However, as shown in Figure~\ref{fig:histogram}c,
          all of our clouds have $A_V<4\magni$ 
          and are mostly diffuse or translucent, not molecular.
          Indeed, for 38 of our 40 sightlines, 
          a single linear relation describes 
          the relation between $\WDIB$ and $E(B-V)$ very well,
          except the $\lambda$6203$\Angstrom$
          and $\lambda$6284$\Angstrom$ DIBs
          which will be discussed below. 
\item If a sightline has been observed only once,
         we will compare the reported EW of a given DIB
         with the $\WDIB$--$E(B-V)$ relation
         derived above. If the reported EW is within 
         3$\sigma$ of the predicted value given by
         the $\WDIB$--$E(B-V)$ relation, we will adopt
         this EW as well as the reported uncertainty 
         (denoted by ``$\surd$'' in 
         Tables~\ref{tab:4428}--\ref{tab:6699}),
         otherwise we will not adopt
         (denoted by ``$\times$'' in 
         Tables~\ref{tab:4428}--\ref{tab:6699}).
         If no uncertainty is given in the literature,
         we assign a 10\% uncertainty.\footnote{%
              For the DIB EW data tabulated in
              Wu (1972), Herbig (1975, 1993, 2000),
              Seab \& Snow (1984), Josafatsson \& Snow (1987),
              Benvenuti \& Porceddu (1989), 
              Kre{\l}owski et al.\ (1999), Megier et al.\ (2001),
              Wszo{\l}ek \& God{\l}owski (2003), and
              Galazutdinov et al.\ (2004), 
              no uncertainties associated with the DIB EWs
              were given in their tables. 
             Herbig (1993), Kre{\l}owski et al.\ (1999)
             and Galazutdinov et al.\ (2004) all assumed
             a 10\% uncertainty.  
             }            
         For illustration, we show in Figure~\ref{fig:4762EWvsEBV}
         that the $\lambda$4762/4763$\Angstrom$ DIB
         has been observed only once 
         in HD\,21291 (Herbig 1975), 
         HD\,190603 (Herbig 1975), 
         and HD\,199478 (Herbig 1975). 
         We do not adopt the reported EWs
         for these three sightlines since they are 
         not within 3$\sigma$ of that given by 
         the $\WDIB$--$E(B-V)$ relation.
\item If the sightline has been observed twice,
          we will take the weighted mean value of 
          the reported EWs if they are close to each other
          within $\simali$15\%,
          otherwise we will first determine 
          the linear relation between
          $W_{\rm DIB}$ and $E(B-V)$ 
          derived above and take the one
          which falls in the $W_{\rm DIB}$\,--\,$E(B-V)$
          linear relation and reject the one which deviates 
          significantly from that relation by $3\sigma$.
          For illustration, we show in Figure~\ref{fig:4762EWvsEBV}
          that the $\lambda$4762/4763$\Angstrom$ DIB
          has been observed twice in HD\,198478
          (Herbig 1975, Thorburn et al.\ 2003).
          We adopt the one measured by  
          Thorburn et al.\ (2003)
          since their EW is within 3$\sigma$ 
          of that given by the $\WDIB$--$E(B-V)$ relation.
          We do not adopt the EW determined 
          by Herbig (1975) since his reported EW
          deviates considerably (by more than $3\sigma$)
          from that predicted from the $\WDIB$--$E(B-V)$ relation.

\item If the sightline has been observed more than twice
         and for the very same DIB there are three or more sets 
         of EWs reported in the literature, 
         we will take the weighted mean value of those 
         reported EWs which are close to each other 
         within $\simali$15\%
         and reject those which deviate significantly 
         from the majorities by $3\sigma$.

%
\item For the $\lambda$6203$\Angstrom$
          and $\lambda$6284$\Angstrom$ DIBs 
          (see Figure~\ref{fig:6203+6284EWvsEBV}), there appears
          to exist two groups of data each of which corresponds
          to a different slope $d\WDIB/dE(B-V)$. 
          For the $\lambda$6203$\Angstrom$ DIB,
          the EWs derived by Herbig (1975), 
          Benvenuti \& Porceddu (1989), 
          and Thorbun et al.\ (2003) are similar
          and larger than that determined by 
          Megier et al.\ (2005),
          Wszolek \& Godlowski (2003), 
          and Hobbs et al.\ (2008, 2009).
          This is probably because the former authors
          integrated over a wider wavelength range.  
          We consider these data separately
          and label these with larger EWs ``6203(1)''
          and these with smaller EWs ``6203(2)''
          (see Tables~\ref{tab:6203(1)},\ref{tab:6203(2)}).
          Similarly, for the $\lambda$6284$\Angstrom$ DIB, 
          Benvenuti \& Porceddu (1989), 
          D\'esert et al.\ (1995), Snow et al.\ (2002),
          Hobbs et al.\ (2008, 2009), Friedman et al.\ (2011),
          and Raimond et al.\ (2012) derived similar $\WDIB$
          values which are larger than that determined by
          Herbig (1975), Seab \& Snow (1984), 
          and Megier et al.\ (2005). 
          Again, we consider these data separately
          and label these with larger EWs ``6284(1)''
          and these with smaller EWs ``6284(2)''
          (see Tables~\ref{tab:6284(1)},\ref{tab:6284(2)}).     
          
%
\end{itemize}
Finally, we intend to favor recent measurements 
over ``ancient'' measurements. 
By ``ancient'' we mean those measurements
made before 2000.
Also, for a given line of sight we ignore those DIBs
for which $W_{\rm DIB}/E(B-V) < 10\mA\magni^{-1}$. 
These DIBs are rather weak and have large
uncertainties (e.g., see Wszolek \& Godlowski 2003).
To this end, 73 of 2260 ($\simali$3.2\%) are 
ignored and not included in the following 
correlation analysis. 
In Tables~\ref{tab:4428}--\ref{tab:6699}
we list the sources and selections 
of the EWs as well as the finally adopted EWs 
for all 40 DIBs.
%
We have two tables for 
the $\lambda$6203$\Angstrom$ DIB and 
the $\lambda$6284$\Angstrom$ DIB each
and one table for each of the remaining 38 DIBs. 
Therefore, in total we have 42 tables. 
%
%

%

Let  $\WDIBp\equiv \WDIB/E(B-V)$
be the normalized DIB EW. 
We examine the correlation between
$\WDIBp$ and $\AFUVint$.
%
With a Pearson correlation coefficient
of $|r|<0.60$ and a Kendall $|\tau|<0.40$, 
we find that most ($\simali$90\%) 
of the DIBs studied here show no correlation
with the FUV extinction. 
For illustration, we show in 
Figures~\ref{fig:DIB4762}a,
\ref{fig:DIB5780}a, \ref{fig:DIB6660}a
the correlation results for 
the $\lambda$4762$\Angstrom$, 
$\lambda$5780$\Angstrom$, 
and $\lambda$6660/6661$\Angstrom$ DIBs. 
Most DIBs are like these three DIBs and are
not related to the FUV extinction. 
In Table~\ref{tab:correlation} we list
the Pearson correlation coefficients $r$
and the Kendall $\tau$ coefficients and
the corresponding significance levels $p$
for the correlation between each DIB
and $\AFUVint$.
Among 40 DIBs, only one DIB 
--- the $\lambda$4501$\Angstrom$ DIB 
(with $r\approx0.84$ and $\tau\approx0.55$, 
$p\approx0.126$) --- 
correlates with the FUV extinction
(see Figure~\ref{fig:DIB4501}a).
Meanwhile, with $r\approx0.52$
and $\tau\approx0.46$, $p\approx0.05$, 
the $\lambda$6090$\Angstrom$ DIB 
appears to show a tendency of correlating 
with the FUV extinction
(see Figure~\ref{fig:DIB6090}a).
It is also found that two (over 40) DIBs,
the $\lambda$4428/4430$\Angstrom$ DIB 
(with $r\approx-0.69$ and 
$\tau\approx-0.44$, $p\approx0.110$;
see Figure~\ref{fig:DIB4428}a)\footnote{%
   If we exclude HD\,149579 
   (T\"ug \& Schmidt-Kaler 1981), 
   the data point at the upper left corner,
   the anti-correlation between the FUV extinction
   and the $\lambda$4428/4430$\Angstrom$ DIB 
   becomes weaker:  $r\approx-0.51$ and 
   $\tau\approx-0.36$, $p\approx0.22$.
   Further spectroscopic observations 
   of this DIB along the line of sight 
   toward HD\,149579 
   will be very helpful.
   }
and the $\lambda$6699$\Angstrom$ DIB 
(with $r\approx-0.61$ and 
$\tau\approx-0.43$, $p\approx0.072$; 
see Figure~\ref{fig:DIB6699}a)
are somewhat negatively correlated 
with the FUV extinction.

\section{Discussion}\label{sec:discussion}
In \S\ref{sec:correlation} we have shown that
the EWs of most of the DIBs show no correlation
with $\AFUVint$, the FUV extinction obtained by
decomposing the observed extinction curve into
three components. As $\AFUVint$ measures 
the column density of the FUV absorber
(see eq.\,\ref{eq:NFUV}) and the DIB EW ($\WDIB$) 
measures the column density of the DIB carrier
(see eq.\,\ref{eq:NDIB}), the lack of correlation between
$\AFUVint$ and $\WDIB$ implies they are not 
from the same carrier. 
If PAHs are responsible for DIBs 
(e.g., see Crawford et al.\ 1985, 
L\'eger \& d'Hendecourt 1985, 
van der Zwet \& Allamandola 1985, 
Salama et al.\ 1999, 2011),
then they cannot be the sole contributor 
to the observed FUV extinction. 
Other dust components
such as small graphitic grains and small silicate grains 
must be present to account for at least part of 
the observed FUV extinction. 
The silicate-graphite-PAH model (WD01, Li \& Draine 2001a)
fits into this scenario. As shown in Figure~\ref{fig:extcurvFit},
the FUV extinction is roughly equally contributed 
by small graphite grains and small silicate grains.
We note that an upper limit on the quantity of 
nano-sized silicate grains can be placed based on
the nondetection of the 9.7$\mum$ emission feature
in the diffuse ISM (Li \& Draine 2001b).

\begin{figure}
\includegraphics[width=16cm,angle=0]{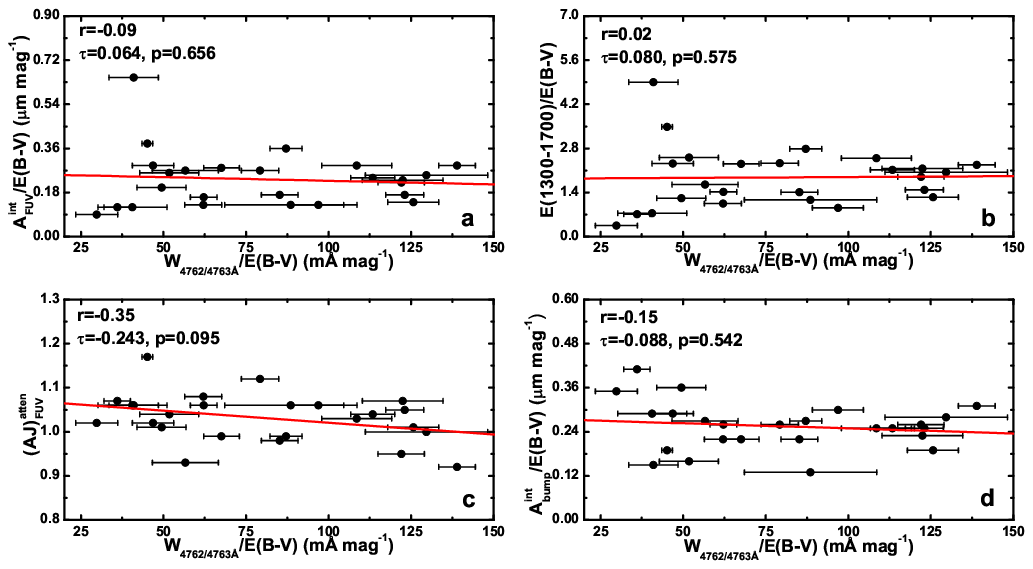} 
\caption{\footnotesize
             \label{fig:DIB4762}
            Correlation diagrams of 
            the $\lambda$4762$\Angstrom$/4763$\Angstrom$ DIB
            with (a) $\AFUVint$, (b) $E(1300-1700)$,
           (c) $\AJFUVatten$, and (d) $\Abumpint$.
           All quantities are normalized to $E(B-V)$. 
           }
\end{figure}

\begin{figure}
\includegraphics[width=16cm,angle=0]{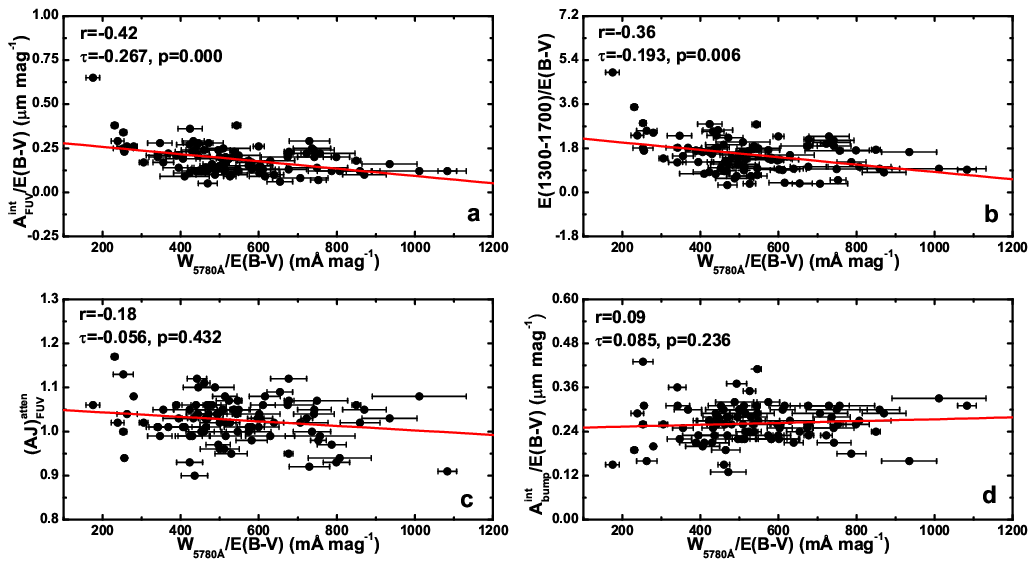} 
\caption{\footnotesize
          \label{fig:DIB5780}
          Same as Figure~\ref{fig:DIB4762}
          but for the $\lambda$5780$\Angstrom$ DIB.
          }
\end{figure}

\begin{figure}
\includegraphics[width=16cm,angle=0]{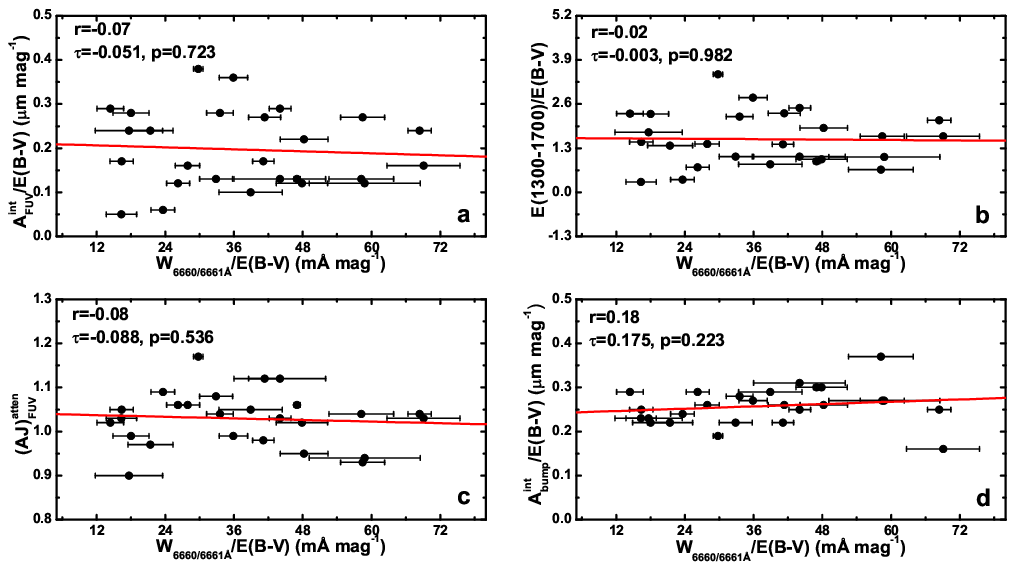} 
\caption{\footnotesize
          \label{fig:DIB6660}
          Same as Figure~\ref{fig:DIB4762}  
          but for the $\lambda$6660$\Angstrom$/6661$\Angstrom$ DIB.
          }
\end{figure}

\begin{figure}
\includegraphics[width=16cm,angle=0]{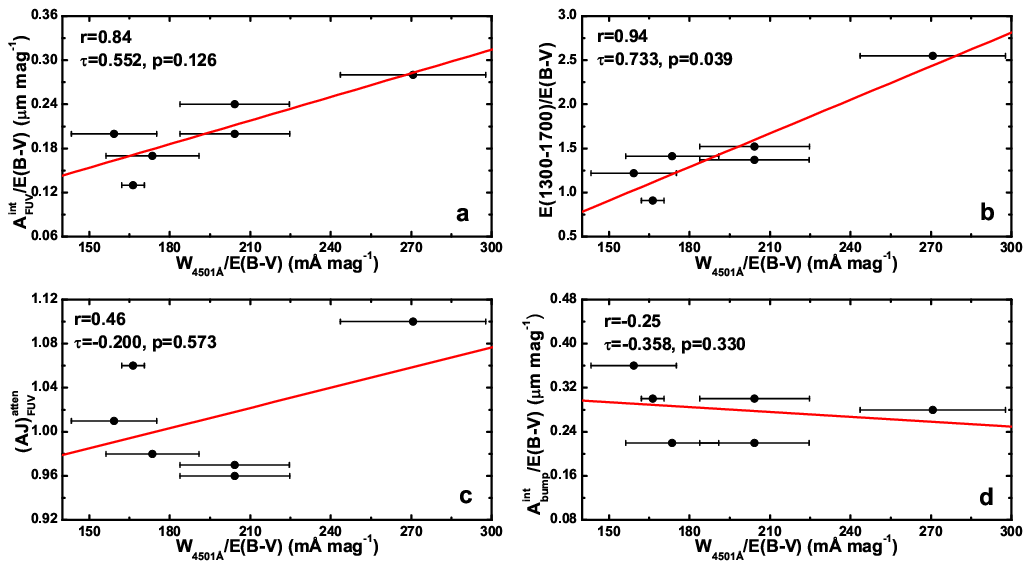} 
\caption{\footnotesize
          \label{fig:DIB4501}
          Same as Figure~\ref{fig:DIB4762} 
          but for the $\lambda$4501$\Angstrom$ DIB.
          }
\end{figure}

\begin{figure}
\includegraphics[width=16cm,angle=0]{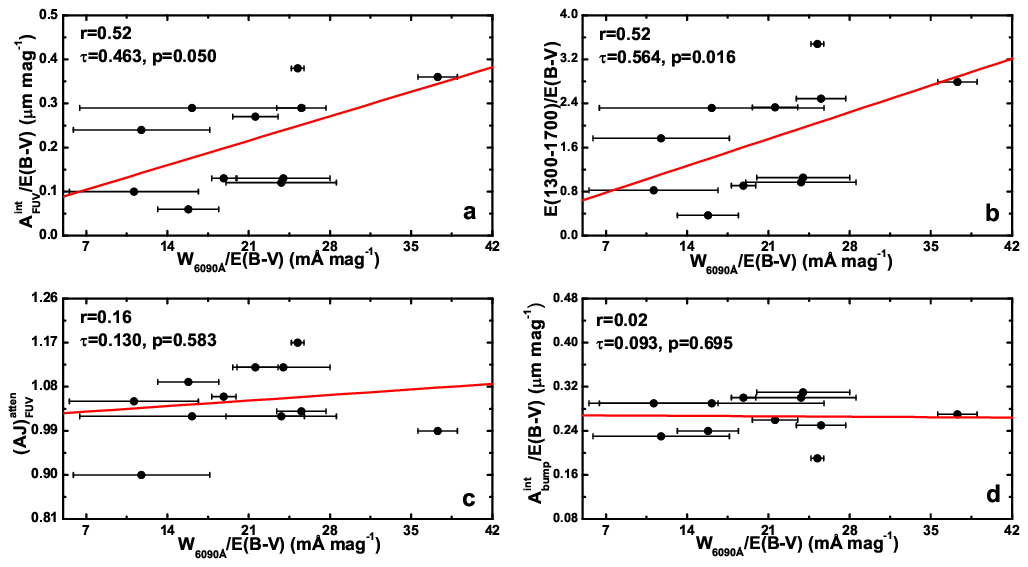} 
\caption{\footnotesize
          \label{fig:DIB6090}
          Same as Figure~\ref{fig:DIB4762} 
          but for the $\lambda$6090$\Angstrom$ DIB.
          }
\end{figure}

\begin{figure}
\includegraphics[width=16cm,angle=0]{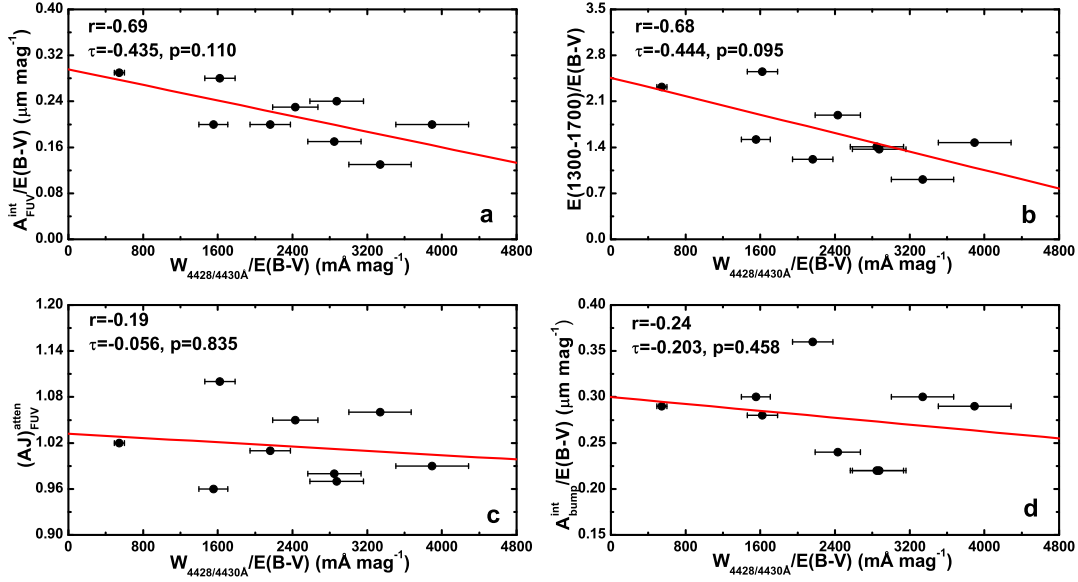} 
\caption{\footnotesize
          \label{fig:DIB4428}
          Same as Figure~\ref{fig:DIB4762} 
          but for the $\lambda$4428/4430$\Angstrom$ DIB.
          }
\end{figure}

\begin{figure}
\includegraphics[width=16cm,angle=0]{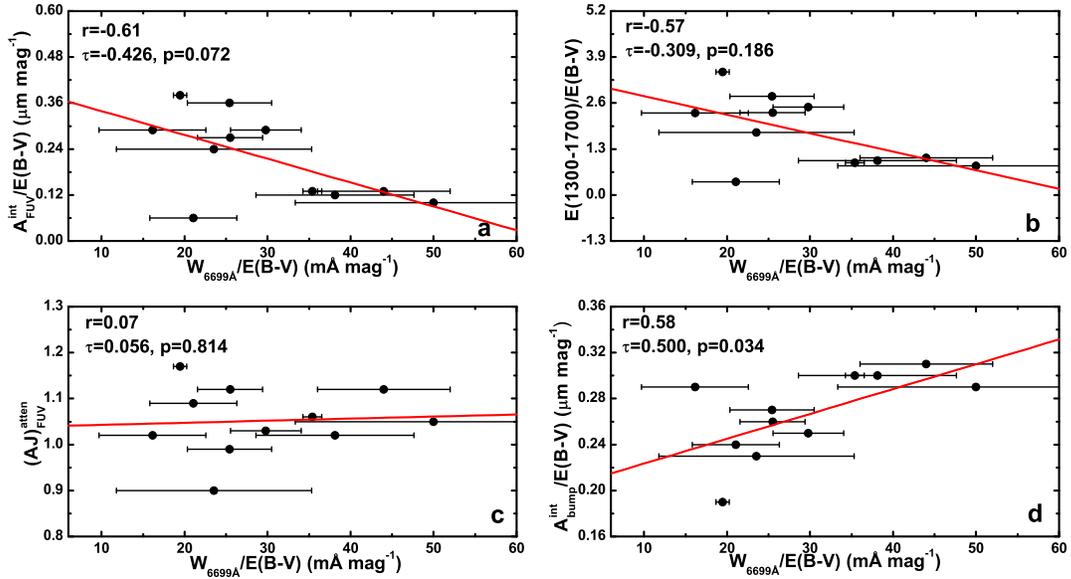} 
\caption{\footnotesize
         \label{fig:DIB6699}
          Same as Figure~\ref{fig:DIB4762} 
          but for the $\lambda$6699$\Angstrom$ DIB. 
          }
\end{figure}

In the literature, the extinction color excess 
$E(\lambda_1-\lambda_2)$ instead of $\AFUVint$
is often used for exploring the correlation 
between DIBs and the FUV extinction 
(see \S\ref{sec:history}).
To be complete, we have also considered 
the correlation of $\WDIB$ with 
$E(1300-1700)\equiv A_{1300}-A_{1700}$, 
the difference between the extinction 
at $\lambda=1300\Angstrom$ and 
$\lambda=1700\Angstrom$.
The correlation results are close to that derived from
$\WDIB$ and $\AFUVint$ 
(see Table~\ref{tab:correlation};
also see Figures~\ref{fig:DIB4762}b--\ref{fig:DIB6699}b
for the selected DIBs).
This suggests that $E(1300-1700)$ is 
a valid indicator of the FUV extinction.
%

To investigate whether and how the DIB carriers are
affected by the interstellar UV radiation, we have also
considered the correlation between $\WDIB$ and $\AJFUVatten$,
with the latter measures the starlight attenuation:
\begin{equation}
\label{eq:AJFUVatten}
\AJFUVatten \equiv \int_{912\Angstrom}^{\infty} 
\JISRF \exp\left\{-\frac{1}{2}\,\left(\frac{A_V}{1.086}\right)
\left(\frac{A_\lambda}{A_V}\right)\right\}\,d\lambda/
\int_{912\Angstrom}^{\infty}  \JISRF\,d\lambda ~~,
\end{equation}
where $\JISRF$ is the Mathis, Mezger \& Panagia (1983)
interstellar radiation field (hereafter MMP83 ISRF),
$A_V$ is the line-of-sight visual extinction,
and $A_\lambda/A_V$ is the line-of-sight extinction curve.
In eq.\,\ref{eq:AJFUVatten}, the factor ``$\frac{1}{2}$'' 
accounts for the fact that $A_V$ is for the whole cloud 
and even the starlight in the cloud center 
only suffers a total amount of 
$1/2\,A_V$ visual extinction.
We find that $\WDIB$ and $\AJFUVatten$ are 
not correlated (see Table~\ref{tab:correlation};
also see Figures~\ref{fig:DIB4762}c--\ref{fig:DIB6699}c
for the selected DIBs).
This suggests that the DIB carriers are rather robust 
and the attenuation of the interstellar UV radiation 
does not necessarily affect their survival in the ISM. 
%
%

As shown in \S\ref{sec:correlation},
while most of 
the DIBs studied here
are not related to the FUV extinction, 
the $\lambda$4501$\Angstrom$ 
and $\lambda$6090$\Angstrom$ DIBs 
somewhat positively correlate with
the FUV extinction measured in terms of $\AFUVint$
and  $E(1300-1700)$, but not with $\AJFUVatten$.
The fact that these two DIBs are not
correlated with $\AJFUVatten$ indicates
that their correlation with the FUV extinction 
is not due to the attenuation of the UV radiation
which could shield their carriers 
from being photodissociated.
Perhaps the carriers of these two DIBs
are related to the carriers of the FUV extinction,
e.g., with the former originating from
the collisional grinding of the latter.
Laboratory experiments have shown that,
induced by shocks (i.e., grain-grain collisions), 
small hydrogenated amorphous carbon (HAC) 
grains in the ISM may decompose into
PAHs and fullerenes (Scott et al.\ 1997).
If small HAC grains are responsible for
the FUV extinction while PAHs and/or 
fullerenes are responsible 
for (some of) the DIBs, it is natural to
expect $\WDIB$ to positively correlate
with $\AFUVint$ and $E(1300-1700)$.
However, it is puzzling why this is only
true for the $\lambda$4501$\Angstrom$ 
and $\lambda$6090$\Angstrom$ DIBs,
not true for the other 38 (of 40) DIBs. 
Moreover, it is also puzzling that
the $\lambda$4428/4430$\Angstrom$ 
and $\lambda$6699$\Angstrom$ DIBs 
show a negative correlation
with $\AFUVint$ and $E(1300-1700)$,
but not with $\AJFUVatten$.
One may speculate that the creation 
of the carriers of DIBs may require
UV photons (e.g., the carriers are possibly
ions and therefore one needs UV photons to
photoionize the carriers, see Witt 2014).
If this is true, one may argue that 
the anti-correlation between these two DIBs 
and the FUV extinction could be due to 
the reduction of UV photons
in regions with a larger $\AFUVint$ 
or $E(1300-1700)$.
However, the fact that these two DIBs are not
correlated with $\AJFUVatten$ challenges 
this hypothesis.
%
%

It would be useful to explore these DIBs in well-studied regions
of which the physical and chemical conditions are known.
In the future work we will examine the effects of the UV extinction
on DIBs in individual, well-studied regions 
such as the HD\,147889 sightline (Ruiterkamp et al.\ 2005)
and the Small Magellanic Cloud wing and bar regions
(Cox et al.\ 2007). 
It is interesting to note that DIBs are present 
in the Large and Small Magellanic Clouds
(LMC/SMC; Ehrenfreund et al.\ 2002, Cox et al.\ 2006, 2007, 
Welty et al.\ 2006) of which the starlight intensities 
and the shapes of the extinction curves 
differ substantially from that of the Milky Way. 
The SMC bar extinction curve
is characterized by a steep, featureless, almost linear rise 
with $\lambda^{-1}$ and lacks the 2175$\Angstrom$ bump
(see Li et al.\ 2005), while the starlight intensity of the SMC bar
is stronger than the Milky Way diffuse ISM by a factor
of $\simali$30 (see Li \& Draine 2002).  
Quantitative studies of DIBs in the LMC and SMC sightlines
would allow us to gain insight into the effects of UV radiation
and dust extinction on DIBs (e.g., see Cox \& Spaans 2006).

Xiang et al.\ (2011) explored the relation between the EWs 
of nine DIBs and the 2175$\Angstrom$ extinction bump of
84 sightlines, using $\pi c_3/2\gamma$ as a measure of
the bump strength. No correlation was found. 
In this work we have also investigated the correlation
between $\Abumpint$, the wavelength-integrated bump
and $\WDIB$ for 40 DIBs along 97 sightlines. 
Similar to Xiang et al.\ (2011), we find that the 2175$\Angstrom$
bump does not correlate with DIBs,
except the $\lambda$6699$\Angstrom$ DIB 
shows a tendency of correlating with 
the 2175$\Angstrom$ bump.
This suggests that it is unlikely 
for the extinction bump and DIBs 
to share a common carrier
(see Figures~\ref{fig:DIB4762}d--\ref{fig:DIB6699}d
for the selected DIBs),
posing a challenge to the hypothesis 
of PAHs being responsible for
both the 2175$\Angstrom$ bump
(Joblin et al.\ 1992, Li \& Draine 2001a, 
Cecchi-Pestellini et al.\ 2008,
Malloci et al.\ 2008, Steglich et al.\ 2010) 
and DIBs (Crawford et al.\ 1985, 
L\'eger \& d'Hendecourt 1985, 
van der Zwet \& Allamandola 1985, 
Salama et al.\ 1999, 2011).

\begin{figure}
\includegraphics[width=16cm,angle=0]{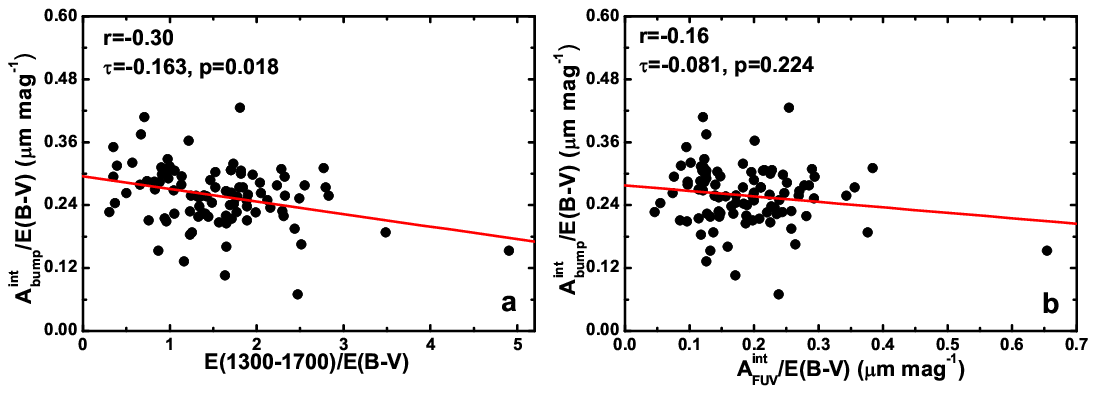} 
\caption{\footnotesize
             \label{fig:FUVBump}
             Correlation diagrams of the 2175$\Angstrom$ 
             extinction bump ($\Abumpint$) 
             with the FUV extinction:
             (a) $\Abumpint$ vs. $E(1300-1700)$;
             (b) $\Abumpint$ vs. $\AFUVint$.   
             }
\end{figure}

It is long known that the 2175$\Angstrom$ extinction bump
does not correlate with the FUV extinction 
(see Greenberg \& Chlewicki 1983). In Figure~\ref{fig:FUVBump}
we plot $\Abumpint$ (see eq.\,\ref{eq:Abumpint}) 
against $E(1300-1700)$ as well as $\Abumpint$ 
against $\AFUVint$, with all quantities normalized to
$E(B-V)$. Our results confirm the earlier findings of
Greenberg \& Chlewicki (1983) that the 2175$\Angstrom$
bump and the FUV extinction are not related,
suggesting that the bump carriers are not a dominant
contributor of the FUV extinction.   

\begin{figure}
\includegraphics[width=16cm,angle=0]{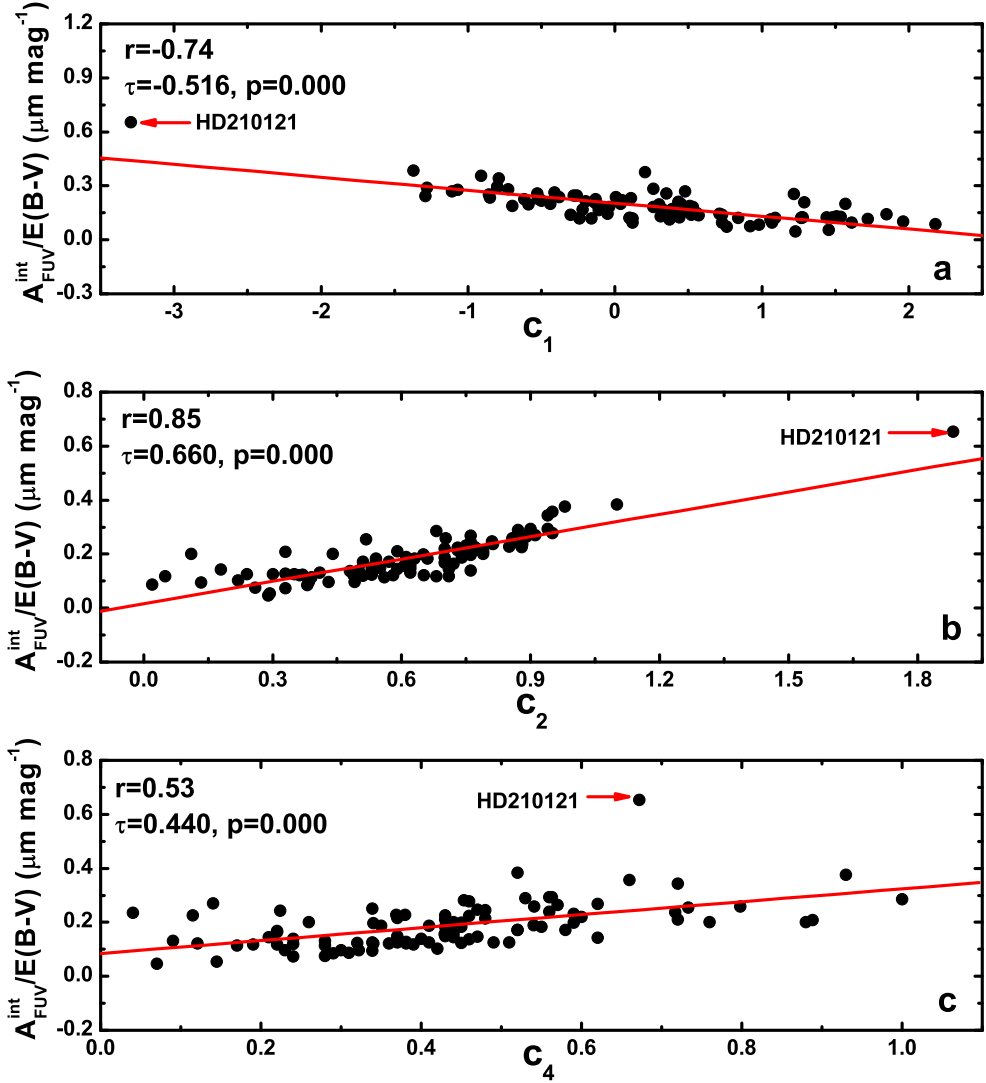} 
\caption{\footnotesize
              \label{fig:FUVvsCj}
             Correlation diagrams of the wavelength-integrated 
             FUV extinction ($\AFUVint$) with the FM parameters
             $c_1$ (a),  $c_2$ (b),  and $c_4$ (c).  
             The extinction curve of the line of sight toward HD\,210121 
             ($R_V\approx 2.1$) is characterzied by a very steep FUV rise
             and a weak 2175$\Angstrom$ bump
             (see Figure~\ref{fig:extcurvFit}, 
             also see Larson et al.\ 1996, Li \& Greenberg 1998). 
             } 
\end{figure}

In the FM parametrization, the FUV extinction is measured
by three parameters: $c_1$, $c_2$ and $c_4$
(see eqs.\,\ref{eq:E2EBV},\,\ref{eq:A2AV}). 
In Figure~\ref{fig:FUVvsCj} we correlate 
the wavelength-integrated FUV extinction ($\AFUVint$) 
with these FM parameters. It is seen that $\AFUVint$
negatively correlates with $c_1$ and positively correlates
with $c_2$. Also, $\AFUVint$ tends to correlate with $c_4$.
Figure~\ref{fig:FUVvsCj} shows that if one really wants to
use one FM parameter to describe the FUV extinction,
$c_2$ would be the most favorable parameter.

\section{Summary}\label{sec:summary}
We have explored the relationship between 
the FUV extinction and 40 DIBs along 97 sightlines. 
The principal results are as follows:
\begin{itemize}
\item We have tried to compile from the literature
          an as large as possible set of sightlines 
          for which both the interstellar extinction curves
          and DIBs have been measured. 
          This leads to a sample of 40 DIBs and 97 sightlines
          which meets the following criterion: 
          (1) in order for our analysis of the correlation
               between DIBs and the UV extinction to be 
               statistically meaningful, we require that,
               for any given DIB, it must have been detected 
               in at least five sightlines, and 
          (2) for any given line of sight, the UV extinction curve 
               must have been measured.
\item We have proposed a decomposition technique 
         to decompose the observed interstellar extinction 
         curve into three Drude-like functions which consist 
        of a visible/near-IR component, a bump peaking
         at 2175$\Angstrom$, and a FUV component.
         This decomposition technique is justified by
          interstellar grain models.
         We argue that the wavelength-integrated FUV
         extinction $\AFUVint$ is the best measure of
         the strength of the FUV extinction.
\item We have compiled the FM extinction parameters and
         the EWs of 40 DIBs along 97 sightlines.
         We have decomposed the extinction curves of these 
         sightlines to obtain $\AFUVint$.
         In the literature, there is quite a rich variety of 
          information on the EWs of DIBs. 
          However, there is no concensus in the DIB
          community on how to derive the DIB EW
          and therefore, it is rather common that different EWs 
          have been reported by different authors 
          for the very same DIB 
          along the very same sightline.
          We have proposed a procedure to  
          carefully select and determine the DIB EWs 
          from the literature, with the original data
          and the selection procedure fully tabulated.
\item We have examined the correlation between the DIB EWs
         and $\AFUVint$. It is found that,
         with a Pearson correlation coefficient
         of $|r|<0.60$ and a Kendall correlation coefficient 
         of $|\tau|<0.40$,
         for most ($\simali$90\%) of the DIBs
         they are not correlated with the FUV extinction.
         We have also studied the relation between DIBs
         and $E(1300-1700)$ and no correlation is found.
         The color excess $E(1300-1700)$ is a quantity 
         often used in the literature to measure the strength 
         of the FUV extinction. We confirm that $E(1300-1700)$ 
         is a valid indicator of the FUV extinction.  
\item It is also found that the wavelength-integrated 2175$\Angstrom$
         bump extinction does not correlate with $\AFUVint$,
         confirming the early findings that the 2175$\Angstrom$ 
         bump and the FUV extinction are not related. 
         We have also examined the relation between $\AFUVint$ 
         and the FM parameters $c_1$, $c_2$ and $c_4$
         (where $c_1$ and $c_2$ are respectively the intercept 
         and slope of the linear extinction component
         while $c_4$ determines the FUV nonlinear extinction rise).
         It is found that the FUV extinction $\AFUVint$ is best
         measured by $c_2$.
\end{itemize}


\acknowledgments{
We thank B.T.~Draine and
the anonymous referee
for very helpful suggestions.
We are supported in part by
NSF AST-1311804,
NNX13AE63G, 
NSFC\,11273022,  NSFC\,U1531108, NSFC\,11473023,
and the University of Missouri Research Board.
}


\begin{deluxetable}{ccccccccc}
\tabletypesize{\tiny}
\tablewidth{400pt}
\tablecaption{\footnotesize
              \label{tab:FMPara}
              Color Excess $E(B-V)$ 
              and the Total-to-Selective Extinction Ratio $\RV$ 
              for Each Sightline
              as well as the FM Parameters 
              Which Fit the Observed Extinction Curve 
              at $\lambda^{-1}>3.3\mum^{-1}$
              for a Given Sightline.
              }
\tablehead{ \colhead{HD/BD} & \colhead{$E(B-V)$}  
            & \colhead{$R_{V}$} & \colhead{$c_{1}$}   
            & \colhead{$c_{2}$}  & \colhead{$c_{3}$}  
            & \colhead{$\gamma$} & \colhead{$x_{0}$} & \colhead{$c_{4}$}
\\
& \colhead{(mag)} & \colhead{} & \colhead{} &  &\colhead{} 
& \colhead{($\mu$m)} & \colhead{($\mu$m)} & \colhead{} 
}
\startdata

     2905  &  0.33$^{(1)}$     &  3.26$^{(2)}$      &    0.39 $^{(3)}$      &   0.62 $^{(3)}$        &    2.04 $^{(3)}$      &  0.79 $^{(3)}$     &   4.579$^{(3)}$    &  0.44 $^{(3)}$      \\  
    15558  &  0.75$^{(4)}$     &  3.09$^{(2)}$      &    0.41 $^{(3)}$      &   0.59 $^{(3)}$        &    3.46 $^{(3)}$      &  0.94 $^{(3)}$     &   4.590$^{(3)}$    &  0.44 $^{(3)}$      \\  
    15570  &  1.02$^{(5)}$     &  3.30$^{(2)}$      &    0.51 $^{(3)}$      &   0.61 $^{(3)}$        &    3.34 $^{(3)}$      &  0.95 $^{(3)}$     &   4.599$^{(3)}$    &  0.54 $^{(3)}$      \\  
    15629  &  0.72$^{(6)}$     &  3.21$^{(2)}$      &   -0.04 $^{(3)}$      &   0.63 $^{(3)}$        &    3.50 $^{(3)}$      &  0.91 $^{(3)}$     &   4.587$^{(3)}$    &  0.43 $^{(3)}$      \\  
    16691  &  0.83$^{(4)}$     &  2.93$^{(7)}$      &   -0.20 $^{(3)}$      &   0.73 $^{(3)}$        &    3.97 $^{(3)}$      &  0.95 $^{(3)}$     &   4.575$^{(3)}$    &  0.48 $^{(3)}$      \\  
    21291  &  0.42$^{(8)}$     &  3.43$^{(2)}$      &    0.26 $^{(3)}$      &   0.68 $^{(3)}$        &    4.98 $^{(3)}$      &  1.07 $^{(3)}$     &   4.539$^{(3)}$    &  1.00 $^{(3)}$      \\  
    21483  &  0.56$^{(1)}$     &  2.90$^{(9)}$      &   -0.41 $^{(9)}$      &   0.89 $^{(9)}$        &    2.85 $^{(9)}$      &  1.10 $^{(9)}$     &   4.62 $^{(9)}$    &  0.57 $^{(9)}$      \\  
    23060  &  0.32$^{(10)}$    &  3.02$^{(2)}$      &    1.96 $^{(3)}$      &   0.22 $^{(3)}$        &    4.17 $^{(3)}$      &  1.06 $^{(3)}$     &   4.557$^{(3)}$    &  0.42 $^{(3)}$      \\
\enddata
\begin{list}{}{}
\item {Data for E(B-V) are taken from $^{(1)}$ Friedman et al.\ (2011); $^{(2)}$ Wegner (2003); 
$^{(3)}$ Jenniskens \& Greenberg (1993);$^{(4)}$ D\'esert et al.\ (1995); 
$^{(5)}$ Sonnentrucher et al.\ (1997);  $^{(6)}$ Krelowski et al.\ (1999); $^{(7)}$Aiello et al.\ (1988); $^{(8)}$ Megier et al.\ (2005); $^{(9)}$ Fitzpatrick \& Massa (2005); $^{(10)}$ Galazutdinov et al.\ (2004); 
 $^{(11)}$ Wittet et al.\ (2004); $^{(12)}$ Fitzpatrick \& Massa (1990); $^{(13)}$ Josafatsson \& Snow (1987); $^{(14)}$ Geminale (2006); $^{(15)}$ Seab \& Snow (1984);  
 $^{(16)}$ Herbig (1993); $^{(17)}$ Weselak et al.\ (2008); $^{(18)}$ Gordon et al (2009); $^{(19)}$ Sofia et al.\ (2005); 
$^{(20)}$ Vos et al.\ (2011); $^{(21)}$ Raimond et al.\ (2012); $^{(22)}$ Lewis et al.\ (2005); $^{(23)}$ Valencic et al.\ (2004);  
$^{(24)}$ Herbig (2000);  $^{(25)}$ Cox et al.\ (2007); $^{(26)}$ Valencic et al.\ (2003); $^{(27)}$ Benvenuti \& Porceddu (1989).  }

\end{list}
\end{deluxetable}


\begin{deluxetable}{ccccccccccccccc}
\tabletypesize{\tiny}
\tablewidth{500pt}
\tablecaption{\footnotesize
              \label{tab:DrudePara}
              Parameters for Fitting the Extinction Curve 
              of Each Sightline with Three Drude-Like Functions. 
              Also Tabulated Are the Integrated Extinction 
              of the Visual Component ($\AVISint$), 
              the Bump ($\Abumpint$), 
              the FUV Component ($\AFUVint$),
              and $\chi^2$, the measure of the goodness
              of the decompositional fit 
              (see eq.\,\ref{eq:chisq}). 
              }
\tablehead{
          \colhead{HD/BD} 
          & \colhead{$a_{1}$}  
          & \colhead{$a_{2}$}  
          & \colhead{$a_{3}$}  
          & \colhead{$a_{4}$}  
          & \colhead{$a_{5}$}
          & \colhead{$a_{6}$} 
          & \colhead{$a_{7}$} 
          & \colhead{$a_{8}$} 
          & \colhead{$\lambdaVIS$} 
          & \colhead{$\lambdaFUV$} 
          & \colhead{$\AVISint$} 
          & \colhead{$\Abumpint$}  
          & \colhead{$\AFUVint$}
          & \colhead{$\chi^{2}$}
\\
\colhead{} & \colhead{} & \colhead{} 
           & \colhead{} & \colhead{}
           & \colhead{} & \colhead{}
           & \colhead{} & \colhead{} 
           & \colhead{($\mu$m)} 
           &\colhead{($\mu$m)} 
           & \colhead{($\mu$m)} 
           & \colhead{($\mu$m)}  
           & \colhead{($\mu$m)}
           & \colhead{}  
}
\startdata
      2905    &   2.0   &  100.0  & 241.775  &  -1.975  &   0.022   &   7.656   &  4.703  &  23.567   &  0.046    &   0.080   &   1.331   &   0.049   &     0.049  &     0.479   \\     
     15558    &   2.0   &   95.0  & 235.507  &  -1.960  &   0.047   &   8.784   &  9.258  &  39.088   &  0.046    &   0.080   &   1.329   &   0.082   &     0.047  &     0.525   \\     
     15570    &   2.0   &  110.0  & 250.737  &  -1.960  &   0.040   &   7.478   &  3.525  &  23.589   &  0.046    &   0.080   &   1.319   &   0.073   &     0.057  &     0.405   \\     
     15629    &   2.0   &  105.0  & 245.173  &  -1.960  &   0.051   &   6.219   &  1.351  &  14.028   &  0.046    &   0.080   &   1.319   &   0.085   &     0.057  &     0.474   \\     
     16691    &   2.0   &   90.0  & 229.627  &  -1.955  &   0.065   &   6.366   &  3.252  &  27.718   &  0.046    &   0.076   &   1.329   &   0.104   &     0.073  &     0.544   \\     
     21291    &   2.0   &  110.0  & 250.553  &  -1.960  &   0.044   &   8.784   & 15.000  & 111.363   &  0.046    &   0.075   &   1.318   &   0.081   &     0.083  &     0.417   \\     
     21483    &   2.0   &   80.0  & 221.249  &  -1.955  &   0.034   &   5.874   &  0.973  &  15.782   &  0.046    &   0.082   &   1.353   &   0.057   &     0.091  &     0.655   \\     
     23060    &   2.0   &  130.0  & 268.767  &  -1.940  &   0.071   &   9.520   & 12.770  &  42.000   &  0.046    &   0.078   &   1.302   &   0.106   &     0.034  &     1.673  \\
\enddata
\end{deluxetable}

\begin{deluxetable}{ccccccc}
\tabletypesize{\tiny}
\tablewidth{400pt}
\tablecaption{\footnotesize
              \label{tab:4428}
              Sources for the Equivalent Widths 
              of the 4428/4430$\Angstrom$ DIB.
              }
\tablehead{ \colhead{HD/BD} & \colhead{$E(B-V)$}  & \colhead{$W_{4428/4430}$} & \colhead{Adopt"$\surd$"}  & \colhead{Sources} & \colhead{$E(B-V)$}  & \colhead{$W_{4428/4430}$(m$\AngstromAL$)} 
\\
\colhead{} & \colhead{(mag)}  & \colhead{(m$\AngstromAL$)} & \colhead{or Reject"$\times$"}  & \colhead{}  & \colhead{This Work}   & \colhead{This Work}   
}
\startdata

21291   &  0.41    &      665.0              &    $\surd$  &   Herbig (1975)                  &  0.41    &      665.0$\pm66.5$              \\    
147889  &  1.08    &     2335.0              &    $\surd$  &   Herbig (1975)                  &  1.08    &     2335.0$\pm233.5$              \\   
149757  &  0.33    &      180.0              &    $\surd$  &   T\"ug \& Schmidt-Kaler (1981)  &  0.33    &      180.0$\pm18.0$              \\    
152236  &  0.65    &     1580.0              &    $\surd$  &   T\"ug \& Schmidt-Kaler (1981)  &  0.65    &     1580.0$\pm158.0$              \\   
152249  &  0.48    &     1870.0              &    $\surd$  &   T\"ug \& Schmidt-Kaler (1981)  &  0.48    &     1870.0$\pm187.0$              \\   
183143  &  1.27    &     4240.0              &    $\surd$  &   T\"ug \& Schmidt-Kaler (1981)  &  1.27    &     4240.0$\pm424.0$              \\   
183143  &  1.27    &     5700.0$\pm$43.3     &    $\times$ &   Hobbs et al.\ (2009)            &          &                                   \\   
190603  &  0.71    &     2040.0              &    $\surd$  &   Herbig (1975)                  &  0.71    &     2040.0$\pm204.0$              \\
\enddata
\end{deluxetable}

\begin{deluxetable}{ccccccc}
\tabletypesize{\tiny}
\tablewidth{400pt}
\tablecaption{\footnotesize
              \label{tab:4501}
              Sources for the Equivalent Widths 
              of the 4501$\Angstrom$ DIB.
              }
\tablehead{ \colhead{HD/BD} & \colhead{$E(B-V)$}  & \colhead{$W_{4501}$} & \colhead{Adopt"$\surd$"}  & \colhead{Sources} & \colhead{$E(B-V)$}  & \colhead{$W_{4501}$(m$\AngstromAL$)} 
\\
\colhead{} & \colhead{(mag)}  & \colhead{(m$\AngstromAL$)} & \colhead{or Reject"$\times$"}  & \colhead{}  & \colhead{This Work}   & \colhead{This Work}   
}
\startdata

  21291   &  0.41    &    111              &  $\surd$   & Herbig (1975)         &   0.41    &  111$\pm11.1$     \\
 147889   &  1.08    &    172              &  $\surd$   & Herbig (1975)         &   1.08    &  172$\pm17.2$     \\
 183143   &  1.27    &    211.2$\pm$5.4    &  $\surd$   & Hobbs et al.\ (2009)   &   1.27    &  211.2$\pm5.4$   \\ 
 190603   &  0.71    &    145              &  $\surd$   & Herbig (1975)         &   0.71    &  145$\pm14.5$     \\
 198478   &  0.53    &     92              &  $\surd$   & Herbig (1975)         &   0.53    &   92$\pm9.2$     \\ 
 199478   &  0.47    &     96              &  $\surd$   & Herbig (1975)         &   0.47    &   96$\pm9.6$     \\ 
 204827   &  1.11    &     31.5$\pm$3.6    &  $\times$  & Hobbs et al.\ (2008)   &           &                  \\ 
\enddata
\end{deluxetable}


\begin{deluxetable}{ccccccc}
\tabletypesize{\tiny}
\tablewidth{400pt}
\tablecaption{\footnotesize
              \label{tab:4726}
              Sources for the Equivalent Widths of the 4726$\Angstrom$ DIB.
              }
\tablehead{ \colhead{HD/BD} & \colhead{$E(B-V)$}  & \colhead{$W_{4726}$} & \colhead{Adopt"$\surd$"}  & \colhead{Sources} & \colhead{$E(B-V)$}  & \colhead{$W_{4726}$(m$\AngstromAL$)} 
\\
\colhead{} & \colhead{(mag)}  & \colhead{(m$\AngstromAL$)} & \colhead{or Reject"$\times$"}  & \colhead{}  & \colhead{This Work}   & \colhead{This Work}   
}
\startdata

 21291  &  0.41  &       63         &  $\surd$  &    Herbig (1975)        &  0.41      &         63$\pm6.3$     \\  
147889  &  1.08  &      354         &  $\surd$  &   Herbig (1975)         &  1.08      &        354$\pm35.4$     \\ 
190603  &  0.71  &      198         &  $\surd$  &   Herbig (1975)         &  0.71      &        198$\pm19.8$     \\ 
198478  &  0.53  &      110         &  $\surd$  &   Herbig (1975)         &  0.53      &        110$\pm11.0$     \\ 
199478  &  0.47  &      130         &  $\surd$  &   Herbig (1975)         &  0.47      &        130$\pm13.0$     \\ 
204827  &  1.11  &  283.7$\pm$4.2   &  $\surd$  &   Hobbs et al.\ (2008)   &  1.11      &        283.7$\pm4.2$   \\  
\enddata
\end{deluxetable}
        
\begin{deluxetable}{ccccccc}
\tabletypesize{\tiny}
\tablewidth{400pt}
\tablecaption{\footnotesize
              \label{tab:4762}
              Sources for the Equivalent Widths of the 4762/4763$\Angstrom$ DIB.
              }
\tablehead{ \colhead{HD/BD} & \colhead{$E(B-V)$}  & \colhead{$W_{4762/4763}$} & \colhead{Adopt"$\surd$"}  & \colhead{Sources} & \colhead{$E(B-V)$}  & \colhead{$W_{4762/4763}$(m$\AngstromAL$)} 
\\
\colhead{} & \colhead{(mag)}  & \colhead{(m$\AngstromAL$)} & \colhead{or Reject"$\times$"}  & \colhead{}  & \colhead{This Work}   & \colhead{This Work}      
}
\startdata

  21291  &   0.41    &   139              &   $\times$    &  Herbig  (1975)          &          &                       \\ 
 21483  &   0.56    &    29$\pm$5        &   $\surd$     &  Thorburn et al.\ (2003)  &   0.56   &  29$\pm$5             \\ 
 30614  &   0.30    &    17$\pm$3        &   $\surd$     &  Thorburn et al.\ (2003)  &   0.30   &  17$\pm$3             \\ 
 34078  &   0.52    &    64$\pm$3        &   $\surd$     &  Thorburn et al.\ (2003)  &   0.52   &  64$\pm$3             \\ 
 37903  &   0.35    &    31$\pm$7        &   $\surd$     &  Thorburn et al.\ (2003)  &   0.35   &  31$\pm$7             \\ 
 41117  &   0.45    &    51$\pm$3        &   $\surd$     &  Thorburn et al.\ (2003)  &   0.45   &  51$\pm$3             \\ 
 42087  &   0.36    &    50$\pm$2        &   $\surd$     &  Thorburn et al.\ (2003)  &   0.36   &  50$\pm$2             \\ 
 46202  &   0.49    &    60$\pm$6        &   $\surd$     &  Thorburn et al.\ (2003)  &   0.49   &  60$\pm$6             \\
\enddata
\end{deluxetable}

\begin{deluxetable}{ccccccc}
\tabletypesize{\tiny}
\tablewidth{400pt}
\tablecaption{\footnotesize
              \label{tab:5487}
              Sources for the Equivalent Widths of the 5487$\Angstrom$ DIB.
              }
\tablehead{ \colhead{HD/BD} & \colhead{$E(B-V)$}  & \colhead{$W_{5487}$} & \colhead{Adopt"$\surd$"}  & \colhead{Sources} & \colhead{$E(B-V)$}  & \colhead{$W_{5487}$(m$\AngstromAL$)} 
\\
\colhead{} & \colhead{(mag)}  & \colhead{(m$\AngstromAL$)} & \colhead{or Reject"$\times$"}  & \colhead{}  & \colhead{This Work}   & \colhead{This Work}         
}
\startdata
 2905      &     0.33        &            69$\pm$3    &  $\surd$    &     Fridman et al.\ (2011)        &         0.32  &   70.16$\pm$5.40      \\  
 2905      &     0.32        &            78          &  $\surd$    &      Herbig (1975)               &               &                       \\  
 21291     &     0.41        &            77           &  $\surd$    &     Herbig (1975)               &         0.41  &   77.00$\pm$7.70      \\  
 21483     &     0.56        &            56$\pm$10   &  $\surd$    &     Fridman et al.\ (2011)        &         0.56  &   56.00$\pm$10.00     \\  
 30614     &     0.30        &           35$\pm$5     &  $\surd$    &       Fridman et al.\ (2011)      &         0.30  &   35.00$\pm$5.00      \\  
 34078     &     0.52        &           48$\pm$5     &  $\surd$    &      Fridman et al.\ (2011)       &         0.52  &   57.91$\pm$7.00      \\  
 34078     &     0.51        &           90           &  $\surd$    &      Herbig (1975)               &               &                       \\  
 37367     &     0.40        &         95$\pm$9       &  $\surd$    &      Fridman et al.\ (2011)       &         0.40  &   95.00$\pm$9.00      \\
\enddata
\end{deluxetable}

\begin{deluxetable}{ccccccc}
\tabletypesize{\tiny}
\tablewidth{400pt}
\tablecaption{\footnotesize
              \label{tab:5544}
              Sources for the Equivalent Widths of the 5544$\Angstrom$ DIB.
              }
\tablehead{ \colhead{HD/BD} & \colhead{$E(B-V)$}  & \colhead{$W_{5544}$} & \colhead{Adopt"$\surd$"}  & \colhead{Sources} & \colhead{$E(B-V)$}  & \colhead{$W_{5544}$(m$\AngstromAL$)} 
\\
\colhead{} & \colhead{(mag)}  & \colhead{(m$\AngstromAL$)} & \colhead{or Reject"$\times$"}  & \colhead{}  & \colhead{This Work}   & \colhead{This Work}   
}
\startdata
 21483   &   0.56       &      12.7$\pm$1    &    $\surd$   &  Thorburn et al.\ (2003)      &        0.56    &    12.70$\pm$1.00     \\
 27778   &   0.37       &      7.5$\pm$1     &    $\surd$   &  Thorburn et al.\ (2003)      &        0.37    &     7.50$\pm$1.00     \\
 30614   &   0.30       &      9.4$\pm$1     &    $\surd$   &  Thorburn et al.\ (2003)      &        0.30    &     9.40$\pm$1.00     \\
 34078   &   0.52       &     13.0$\pm$2     &    $\surd$   &  Thorburn et al.\ (2003)      &        0.52    &    13.00$\pm$2.00     \\
 37903   &   0.35       &      4.1$\pm$1     &    $\surd$   &  Thorburn et al.\ (2003)      &        0.35    &     4.10$\pm$1.00     \\
 41117   &   0.45       &     16.0$\pm$1     &    $\surd$   &  Thorburn et al.\ (2003)      &        0.45    &    16.00$\pm$1.00     \\
 42087   &   0.36       &     16.0$\pm$2     &    $\surd$   &  Thorburn et al.\ (2003)      &        0.36    &    16.00$\pm$2.00     \\
 46202   &   0.49       &     18.7$\pm$1.5   &    $\surd$   &  Thorburn et al.\ (2003)      &        0.49    &    18.70$\pm$1.50     \\
\enddata
\end{deluxetable}

\begin{deluxetable}{ccccccc}
\tabletypesize{\tiny}
\tablewidth{400pt}
\tablecaption{\footnotesize
              \label{tab:5705}
              Sources for the Equivalent Widths of the 5705$\Angstrom$ DIB.
              }
\tablehead{ \colhead{HD/BD} & \colhead{$E(B-V)$}  & \colhead{$W_{5705}$} & \colhead{Adopt"$\surd$"}  & \colhead{Sources} & \colhead{$E(B-V)$}  & \colhead{$W_{5705}$(m$\AngstromAL$)} 
\\
\colhead{} & \colhead{(mag)}  & \colhead{(m$\AngstromAL$)} & \colhead{or Reject"$\times$"}  & \colhead{}  & \colhead{This Work}   & \colhead{This Work}   
}
\startdata

2905       &     0.33            &           66$\pm$7          &  $\surd$     &      Friedman et al.\ (2011)             &     0.32     &    58.48$\pm$6.20     \\  
2905       &     0.32            &           54                &  $\surd$     &      Herbig (1975)                      &              &                        \\ 
21291      &     0.42            &           43                &  $\surd$     &      Josafatsson and Snow (1987)        &     0.42     &    43.00$\pm$4.30      \\ 
21291      &     0.41            &          111                &  $\times$     &      Herbig (1975)                     &              &                        \\ 
21483      &     0.56            &           35$\pm$10         &  $\surd$     &        Friedman et al.\ (2011)           &     0.56     &    35.00$\pm$10.00     \\ 
27778      &     0.37            &           17$\pm$3          &  $\surd$     &       Friedman et al.\ (2011)            &     0.37     &    17.00$\pm$3.00      \\ 
30614      &     0.29            &           33                &  $\surd$     &      Josafatsson and Snow (1987)        &     0.29     &    33.00$\pm$3.30      \\ 
30614      &     0.30            &           54$\pm$10         &  $\times$     &        Friedman et al.\ (2011)          &              &                        \\
\enddata
\end{deluxetable}

\begin{deluxetable}{ccccccc}
\tabletypesize{\tiny}
\tablewidth{400pt}
\tablecaption{\footnotesize
              \label{tab:5707}
              Sources for the Equivalent Widths of the 5707$\Angstrom$ DIB.
              }
\tablehead{ \colhead{HD/BD} & \colhead{$E(B-V)$}  & \colhead{$W_{5707}$} & \colhead{Adopt"$\surd$"}  & \colhead{Sources} & \colhead{$E(B-V)$}  & \colhead{$W_{5707}$(m$\AngstromAL$)} 
\\
\colhead{} & \colhead{(mag)}  & \colhead{(m$\AngstromAL$)} & \colhead{or Reject"$\times$"}  & \colhead{}  & \colhead{This Work}   & \colhead{This Work}   
}
\startdata

15558     &  0.75 	 	  & 146.3$\pm$8.3      &  $\surd$   &     D\'esert et al.\ (1995)   &   0.75 	     &  146.3$\pm$8.3     \\  
16691     &  0.83 	 	  & 184.3$\pm$16.6     &  $\surd$   &     D\'esert et al.\ (1995)   &   0.83 	     &  184.3$\pm$16.6    \\  
23060     &  0.34 	 	  & 35.0$\pm$1.4       &  $\surd$   &     D\'esert et al.\ (1995)   &   0.34 	     &  35.0$\pm$1.4      \\  
30614     &  0.32 	 	  & 16.0$\pm$2.6       &  $\surd$   &     D\'esert et al.\ (1995)   &   0.32 	     &  16.0$\pm$2.6      \\  
+31643	  &  0.85 	 	  & 113.9$\pm$5.1      &  $\surd$   &     D\'esert et al.\ (1995)   &   0.85 	     &  113.9$\pm$5.1    \\   
34078     &  0.52 	 	  &  60.8$\pm$7.3      &  $\surd$   &     D\'esert et al.\ (1995)   &   0.52 	     &   60.8$\pm$7.3     \\  
36879     &  0.50 	 	  &  95.0$\pm$11.5     &  $\surd$   &     D\'esert et al.\ (1995)   &   0.50 	     &   95.0$\pm$11.5    \\  
37367     &  0.43 	 	  & 131.2$\pm$18.5     &  $\surd$   &     D\'esert et al.\ (1995)   &   0.43 	     &  131.2$\pm$18.5    \\
\enddata
\end{deluxetable}

\begin{deluxetable}{ccccccc}
\tabletypesize{\tiny}
\tablewidth{400pt}
\tablecaption{\footnotesize
              \label{tab:5763}
              Sources for the Equivalent Widths of the 5763$\Angstrom$ DIB.
              }
\tablehead{ \colhead{HD/BD} & \colhead{$E(B-V)$}  & \colhead{$W_{5763}$} & \colhead{Adopt"$\surd$"}  & \colhead{Sources} & \colhead{$E(B-V)$}  & \colhead{$W_{5763}$(m$\AngstromAL$)} 
\\
\colhead{} & \colhead{(mag)}  & \colhead{(m$\AngstromAL$)} & \colhead{or Reject"$\times$"}  & \colhead{}  & \colhead{This Work}   & \colhead{This Work}       
}
\startdata

  2905  &   0.33     &      2                &  $\times$   &     Wszolek \& Godlowski (2003) &           &             \\
 21291  &   0.42     &      4                &  $\times$   &     Wszolek \& Godlowski (2003) &           &             \\
 27778  &   0.35     &      5                &  $\surd$    &    Wszolek \& Godlowski (2003)  &    0.35   &        5$\pm$0.5    \\
 34078  &   0.52     &      2                &  $\times$   &     Wszolek \& Godlowski (2003) &           &             \\
 41117  &   0.45     &      7                &  $\surd$    &    Wszolek \& Godlowski (2003)  &    0.45   &        7$\pm$0.7    \\
 42087  &   0.36     &      5                &  $\surd$    &    Wszolek \& Godlowski (2003)  &    0.36   &        5$\pm$0.5    \\
 47129  &   0.34     &      3                &  $\times$   &     Wszolek \& Godlowski (2003) &           &             \\
143275  &   0.17     &      1                &  $\times$   &     Wszolek \& Godlowski (2003) &           &             \\
\enddata
\end{deluxetable}

\begin{deluxetable}{ccccccc}
\tabletypesize{\tiny}
\tablewidth{400pt}
\tablecaption{\footnotesize
              \label{tab:5766}
              Sources for the Equivalent Widths of the 5766$\Angstrom$ DIB.
              }
\tablehead{ \colhead{HD/BD} & \colhead{$E(B-V)$}  & \colhead{$W_{5766}$} & \colhead{Adopt"$\surd$"}  & \colhead{Sources} & \colhead{$E(B-V)$}  & \colhead{$W_{5766}$(m$\AngstromAL$)} 
\\
\colhead{} & \colhead{(mag)}  & \colhead{(m$\AngstromAL$)} & \colhead{or Reject"$\times$"}  & \colhead{}  & \colhead{This Work}   & \colhead{This Work}        
}
\startdata

   2905  &   0.33  &    9              &   $\surd$  &   Wszolek \& Godlowski (2003)  &        0.33  &    9$\pm$0.9       \\
  21291  &   0.42  &    8              &   $\surd$  &   Wszolek \& Godlowski (2003)  &        0.42  &    8$\pm$0.8       \\
  21483  &   0.56  &   13              &   $\surd$  &   Wszolek \& Godlowski (2003)  &        0.56  &   13$\pm$1.3       \\
  27778  &   0.35  &    6              &   $\surd$  &   Wszolek \& Godlowski (2003)  &        0.35  &    6$\pm$0.6       \\
  34078  &   0.52  &    5              &   $\times$ &   Wszolek \& Godlowski (2003)  &              &                    \\
  41117  &   0.45  &    5              &   $\surd$  &   Wszolek \& Godlowski (2003)  &        0.45  &    5$\pm$0.5       \\
 143275  &   0.17  &    2              &   $\surd$  &   Wszolek \& Godlowski (2003)  &        0.17  &    2$\pm$0.2       \\
 144217  &   0.18  &    3              &   $\surd$  &   Wszolek \& Godlowski (2003)  &        0.18  &    3$\pm$0.3       \\
\enddata
\end{deluxetable}

\begin{deluxetable}{ccccccc}
\tabletypesize{\tiny}
\tablewidth{400pt}
\tablecaption{\footnotesize
              \label{tab:5773}
              Sources for the Equivalent Widths of the 5773$\Angstrom$ DIB.
              }
\tablehead{ \colhead{HD/BD} & \colhead{$E(B-V)$}  & \colhead{$W_{5773}$} & \colhead{Adopt"$\surd$"}  & \colhead{Sources} & \colhead{$E(B-V)$}  & \colhead{$W_{5773}$(m$\AngstromAL$)} 
\\
\colhead{} & \colhead{(mag)}  & \colhead{(m$\AngstromAL$)} & \colhead{or Reject"$\times$"}  & \colhead{}  & \colhead{This Work}   & \colhead{This Work}   
}
\startdata

  2905   &     0.33   &      7             &      $\surd$    &      Wszolek \& Godlowski (2003)     &       0.33   &      7$\pm$0.7      \\ 
 21291   &     0.42   &      5             &      $\surd$    &      Wszolek \& Godlowski (2003)     &       0.42   &      5$\pm$0.5      \\ 
 27778   &     0.35   &      4             &      $\surd$    &      Wszolek \& Godlowski (2003)     &       0.35   &      4$\pm$0.4      \\ 
 41117   &     0.45   &     14             &      $\surd$    &      Wszolek \& Godlowski (2003)     &       0.45   &     14$\pm$1.4      \\ 
 47129   &     0.34   &      6             &      $\surd$    &      Wszolek \& Godlowski (2003)     &       0.34   &      6$\pm$0.6      \\ 
143275   &     0.17   &      3             &      $\surd$    &      Wszolek \& Godlowski (2003)     &       0.17   &      3$\pm$0.3      \\ 
144217   &     0.18   &      3             &      $\surd$    &      Wszolek \& Godlowski (2003)     &       0.18   &      3$\pm$0.3      \\ 
144470   &     0.21   &      6             &      $\surd$    &      Wszolek \& Godlowski (2003)     &       0.21   &      6$\pm$0.6      \\
\enddata
\end{deluxetable}

\begin{deluxetable}{ccccccc}
\tabletypesize{\tiny}
\tablewidth{400pt}
\tablecaption{\footnotesize
              \label{tab:5776}
              Sources for the Equivalent Widths of the 5776$\Angstrom$ DIB.
              }
\tablehead{ \colhead{HD/BD} & \colhead{$E(B-V)$}  & \colhead{$W_{5776}$} & \colhead{Adopt"$\surd$"}  & \colhead{Sources} & \colhead{$E(B-V)$}  & \colhead{$W_{5776}$(m$\AngstromAL$)} 
\\
\colhead{} & \colhead{(mag)}  & \colhead{(m$\AngstromAL$)} & \colhead{or Reject"$\times$"}  & \colhead{}  & \colhead{This Work}   & \colhead{This Work}   
}
\startdata

  2905   &   0.33   &      7             &   $\surd$     &   Wszolek \& Godlowski (2003)     &   0.33   &      7$\pm$0.7        \\  
 21291   &   0.42   &      5             &   $\surd$     &   Wszolek \& Godlowski (2003)     &   0.42   &      5$\pm$0.5        \\  
 34078   &   0.52   &      8             &   $\surd$     &   Wszolek \& Godlowski (2003)     &   0.52   &      8$\pm$0.8        \\  
 41117   &   0.45   &      9             &   $\surd$     &   Wszolek \& Godlowski (2003)     &   0.45   &      9$\pm$0.9        \\  
 42087   &   0.36   &      7             &   $\surd$     &   Wszolek \& Godlowski (2003)     &   0.36   &      7$\pm$0.7        \\  
 47129   &   0.34   &      4             &   $\surd$     &   Wszolek \& Godlowski (2003)     &   0.34   &      4$\pm$0.4       \\   
143275   &   0.17   &      2             &   $\surd$     &   Wszolek \& Godlowski (2003)     &   0.17   &      2$\pm$0.2        \\  
144217   &   0.18   &      4             &   $\surd$     &   Wszolek \& Godlowski (2003)     &   0.18   &      4$\pm$0.4        \\
\enddata
\end{deluxetable}

\begin{deluxetable}{ccccccc}
\tabletypesize{\tiny}
\tablewidth{400pt}
\tablecaption{\footnotesize
              \label{tab:5778}
              Sources for the Equivalent Widths of the 5778$\Angstrom$ DIB.
              }
\tablehead{ \colhead{HD/BD} & \colhead{$E(B-V)$}  & \colhead{$W_{5778}$} & \colhead{Adopt"$\surd$"}  & \colhead{Sources} & \colhead{$E(B-V)$}  & \colhead{$W_{5778}$(m$\AngstromAL$)} 
\\
\colhead{} & \colhead{(mag)}  & \colhead{(m$\AngstromAL$)} & \colhead{or Reject"$\times$"}  & \colhead{}  & \colhead{This Work}   & \colhead{This Work}   
}
\startdata
 2905       &  0.32    &   335    &  $\surd$     &    Herbig(1975)                 &      0.32  &   335.00$\pm$33.50  \\
 21291      &  0.42    &   200    &  $\surd$     &    Josafatsson \& Snow(1987)    &      0.42  &   200.00$\pm$20.00  \\
 21291      &  0.41    &   360    &  $\times$    &    Herbig(1975)                 &            &                     \\
+31643      &  0.77    &   216    &  $\times$    &    Herbig(1975)                 &            &                     \\
 34078      &  0.52    &   305    &  $\surd$     &    Herbig(1975)                 &      0.52  &   305.00$\pm$30.50  \\
 37903      &  0.36    &   170    &  $\surd$     &    Josafatsson \& Snow(1987)    &      0.36  &   170.00$\pm$17.00  \\
 38087      &  0.24    &   150    &  $\surd$     &    Josafatsson \& Snow(1987)    &      0.24  &   150.00$\pm$15.00  \\
 53974      &  0.31    &   180    &  $\surd$     &    Josafatsson \& Snow(1987)    &      0.31  &   180.00$\pm$18.00  \\
\enddata
\end{deluxetable}

\begin{deluxetable}{ccccccc}
\tabletypesize{\tiny}
\tablewidth{400pt}
\tablecaption{\footnotesize
              \label{tab:5780}
              Sources for the Equivalent Widths of the 5780$\Angstrom$ DIB.
              }
\tablehead{ \colhead{HD/BD} & \colhead{$E(B-V)$}  & \colhead{$W_{5780}$} & \colhead{Adopt"$\surd$"}  & \colhead{Sources} & \colhead{$E(B-V)$}  & \colhead{$W_{5780}$(m$\AngstromAL$)} 
\\
\colhead{} & \colhead{(mag)}  & \colhead{(m$\AngstromAL$)} & \colhead{or Reject"$\times$"}  & \colhead{}  & \colhead{This Work}   & \colhead{This Work}        
}
\startdata
2905     &    0.33    &       301               &    $\surd$     &       Wszolek \& Godlowski (2003)           &         0.33  &  308.53$\pm$23.30       \\        
2905     &    0.32    &       288               &    $\surd$     &       Herbig (1975)                        &               &                         \\        
2905     &    0.33    &       314$\pm$5         &    $\surd$     &        Friedman et al.\ (2011)              &               &                         \\        
2905     &    0.33    &       251.7$\pm$4.3     &    $\times$    &       Megier et al.\ (2005)                 &               &                         \\        
2905     &    0.33    &       282               &    $\surd$     &       Herbig (1993)                        &               &                         \\        
2905     &    0.33    &       304$\pm$10        &    $\surd$     &       Snow et al.\ (2002)                   &               &                         \\        
2905     &    0.33    &       279.3             &    $\surd$     &       Krelowski et al.\ (1999)              &               &                         \\        
2905     &    0.33    &       294               &    $\surd$     &       Megier et al.\ (2001)                 &               &                         \\
\enddata
\end{deluxetable}

\begin{deluxetable}{ccccccc}
\tabletypesize{\tiny}
\tablewidth{400pt}
\tablecaption{\footnotesize
              \label{tab:5793}
              Sources for the Equivalent Widths of the 5793$\Angstrom$ DIB.
              }
\tablehead{ \colhead{HD/BD} & \colhead{$E(B-V)$}  & \colhead{$W_{5793}$} & \colhead{Adopt"$\surd$"}  & \colhead{Sources} & \colhead{$E(B-V)$}  & \colhead{$W_{5793}$(m$\AngstromAL$)} 
\\
\colhead{} & \colhead{(mag)}  & \colhead{(m$\AngstromAL$)} & \colhead{or Reject"$\times$"}  & \colhead{}  & \colhead{This Work}   & \colhead{This Work}   
}
\startdata

   2905   &    0.33   &        3            &      $\times$      &      Wszolek \& Godlowski (2003)    &             &            \\  
  21291   &    0.42   &        5            &      $\surd$       &      Wszolek \& Godlowski (2003)    &     0.42    &        5$\pm$0.5   \\  
  21483   &    0.56   &        7            &      $\surd$       &      Wszolek \& Godlowski (2003)    &     0.56    &        7$\pm$0.7      \\  
  27778   &    0.35   &        5            &      $\surd$       &      Wszolek \& Godlowski (2003)    &     0.35    &        5$\pm$0.5      \\  
  34078   &    0.52   &        5            &      $\times$      &      Wszolek \& Godlowski (2003)    &             &            \\  
  41117   &    0.45   &        8            &      $\surd$       &      Wszolek \& Godlowski (2003)    &     0.45    &        8$\pm$0.8      \\  
  42087   &    0.36   &        7            &      $\surd$       &      Wszolek \& Godlowski (2003)    &     0.36    &        7$\pm$0.7      \\  
  47129   &    0.34   &        3            &      $\times$      &      Wszolek \& Godlowski (2003)    &             &            \\
\enddata
\end{deluxetable}

\begin{deluxetable}{ccccccc}
\tabletypesize{\tiny}
\tablewidth{400pt}
\tablecaption{\footnotesize
              \label{tab:5795}
              Sources for the Equivalent Widths of the 5795$\Angstrom$ DIB.
              }
\tablehead{ \colhead{HD/BD} & \colhead{$E(B-V)$}  & \colhead{$W_{5795}$} & \colhead{Adopt"$\surd$"}  & \colhead{Sources} & \colhead{$E(B-V)$}  & \colhead{$W_{5795}$(m$\AngstromAL$)} 
\\
\colhead{} & \colhead{(mag)}  & \colhead{(m$\AngstromAL$)} & \colhead{or Reject"$\times$"}  & \colhead{}  & \colhead{This Work}   & \colhead{This Work}   
}
\startdata

   2905 &    0.33   &       5           &     $\surd$   &   Wszolek \& Godlowski (2003)    &    0.33     &    5$\pm$0.5          \\ 
  21291 &    0.42   &       5           &     $\surd$   &   Wszolek \& Godlowski (2003)    &    0.42     &    5$\pm$0.5          \\ 
  21483 &    0.56   &       9           &     $\surd$   &   Wszolek \& Godlowski (2003)    &    0.56     &    9$\pm$0.9          \\ 
  27778 &    0.35   &       3           &     $\times$  &   Wszolek \& Godlowski (2003)    &             &            \\ 
  34078 &    0.52   &       6           &     $\surd$   &   Wszolek \& Godlowski (2003)    &    0.52     &    6$\pm$0.6          \\ 
  41117 &    0.45   &       6           &     $\surd$   &   Wszolek \& Godlowski (2003)    &    0.45     &    6$\pm$0.6          \\ 
  42087 &    0.36   &       5           &     $\surd$   &   Wszolek \& Godlowski (2003)    &    0.36     &    5$\pm$0.5          \\ 
 143275 &    0.17   &       2           &     $\surd$   &   Wszolek \& Godlowski (2003)    &    0.17     &    2$\pm$0.2          \\
\enddata
\end{deluxetable}

\begin{deluxetable}{ccccccc}
\tabletypesize{\tiny}
\tablewidth{400pt}
\tablecaption{\footnotesize
              \label{tab:5797}
              Sources for the Equivalent Widths of the 5797$\Angstrom$ DIB.
              }
\tablehead{ \colhead{HD/BD} & \colhead{$E(B-V)$}  & \colhead{$W_{5797}$} & \colhead{Adopt"$\surd$"}  & \colhead{Sources} & \colhead{$E(B-V)$}  & \colhead{$W_{5797}$(m$\AngstromAL$)} 
\\
\colhead{} & \colhead{(mag)}  & \colhead{(m$\AngstromAL$)} & \colhead{or Reject"$\times$"}  & \colhead{}  & \colhead{This Work}   & \colhead{This Work}   
}
\startdata

  2905      &      0.33       &            68.9$\pm$2.1      &       $\surd$      &       Megier et al.\ (2005)              &    0.33   &     70.01$\pm$5.42       \\            
  2905      &      0.32       &             114              &       $\times$     &       Herbig (1975)                     &           &                 \\                     
  2905      &      0.33       &            74$\pm$4          &       $\surd$      &       Snow et al.\ (2002)                &           &                 \\                     
  2905      &      0.33       &            72                &       $\surd$      &       Herbig (1993)                     &           &                 \\                     
  2905      &      0.33       &            88                &       $\times$     &       Wszolek \& Godlowski (2003)        &           &                 \\                     
  2905      &      0.33       &            110$\pm$5         &       $\times$     &       Friedman et al.\ (2011)            &           &                 \\                     
  2905      &      0.33       &            69.3              &       $\surd$      &       Krelowski et al.\ (1999)           &           &                 \\                     
  2905      &      0.33       &            69                &       $\surd$      &       Megier et al.\ (2001)              &           &                 \\
\enddata
\end{deluxetable}

\begin{deluxetable}{ccccccc}
\tabletypesize{\tiny}
\tablewidth{400pt}
\tablecaption{\footnotesize
              \label{tab:5809}
              Sources for the Equivalent Widths of the 5809$\Angstrom$ DIB.
              }
\tablehead{ \colhead{HD/BD} & \colhead{$E(B-V)$}  & \colhead{$W_{5809}$} & \colhead{Adopt"$\surd$"}  & \colhead{Sources} & \colhead{$E(B-V)$}  & \colhead{$W_{5809}$(m$\AngstromAL$)} 
\\
\colhead{} & \colhead{(mag)}  & \colhead{(m$\AngstromAL$)} & \colhead{or Reject"$\times$"}  & \colhead{}  & \colhead{This Work}   & \colhead{This Work}   
}
\startdata

  2905  &  0.33  &    16.1$\pm$4.5    &   $\times$    &  Weselak et al.\ (2001)              &               &             \\ 
  2905  &  0.33  &        3           &   $\times$   &  Wszolek \& Godlowski (2003)        &                &              \\          
 21291  &  0.42  &    3.7$\pm$1.6     &   $\times$   &  Weselak et al.\ (2001)              &                &              \\          
 21291  &  0.42  &        4           &   $\times$   &  Wszolek \& Godlowski (2003)        &                &              \\          
 21483  &  0.56  &        4           &   $\times$   &  Wszolek \& Godlowski (2003)        &                &              \\          
 27778  &  0.35  &        1           &   $\times$   &  Wszolek \& Godlowski (2003)        &                &              \\          
 34078  &  0.52  &       7            &   $\surd$    &  Wszolek \& Godlowski (2003)        &        0.52    &    7.00$\pm$0.70      \\ 
 41117  &  0.44  &  13.4$\pm$4.2      &   $\times$   &  Weselak et al.\ (2001)              &        0.45    &    7.00$\pm$0.70      \\
\enddata
\end{deluxetable}

\begin{deluxetable}{ccccccc}
\tabletypesize{\tiny}
\tablewidth{400pt}
\tablecaption{\footnotesize
              \label{tab:5819}
              Sources for the Equivalent Widths of the 5819$\Angstrom$ DIB.
              }
\tablehead{ \colhead{HD/BD} & \colhead{$E(B-V)$}  & \colhead{$W_{5819}$} & \colhead{Adopt"$\surd$"}  & \colhead{Sources} & \colhead{$E(B-V)$}  & \colhead{$W_{5819}$(m$\AngstromAL$)} 
\\
\colhead{} & \colhead{(mag)}  & \colhead{(m$\AngstromAL$)} & \colhead{or Reject"$\times$"}  & \colhead{}  & \colhead{This Work}   & \colhead{This Work}   
}
\startdata

 21291  &   0.42  &    10           &   $\surd$   &   Wszolek \& Godlowski (2003)   &   0.42    &   10$\pm$1.0     \\   
 41117  &   0.45  &    5            &   $\surd$   &   Wszolek \& Godlowski (2003)   &   0.45    &    5$\pm$0.5     \\   
144217  &   0.18  &    2            &   $\surd$   &   Wszolek \& Godlowski (2003)   &   0.18    &    2$\pm$0.2     \\   
144470  &   0.21  &    3            &   $\surd$   &   Wszolek \& Godlowski (2003)   &   0.21    &    3$\pm$0.3     \\   
183143  &   1.27  &   15            &   $\surd$   &   Wszolek \& Godlowski (2003)   &   1.27    &   15$\pm$1.5     \\   
183143  &   1.27  &    7.7$\pm$0.7  &   $\times$  &   Hobbs et al.\ (2009)           &           &          \\   
193322  &   0.41  &    2            &   $\times$  &   Wszolek \& Godlowski (2003)   &           &          \\   
204827  &   1.11  &   9.6$\pm$0.9   &   $\times$  &   Hobbs et al.\ (2008)           &           &          \\
\enddata
\end{deluxetable}

\begin{deluxetable}{ccccccc}
\tabletypesize{\tiny}
\tablewidth{400pt}
\tablecaption{\footnotesize
              \label{tab:5829}
              Sources for the Equivalent Widths of the 5829$\Angstrom$ DIB.
              }
\tablehead{ \colhead{HD/BD} & \colhead{$E(B-V)$}  & \colhead{$W_{5829}$} & \colhead{Adopt"$\surd$"}  & \colhead{Sources} & \colhead{$E(B-V)$}  & \colhead{$W_{5829}$(m$\AngstromAL$)} 
\\
\colhead{} & \colhead{(mag)}  & \colhead{(m$\AngstromAL$)} & \colhead{or Reject"$\times$"}  & \colhead{}  & \colhead{This Work}   & \colhead{This Work}   
}
\startdata

  21291   &   0.42    &     6            &    $\surd$   &    Wszolek \& Godlowski (2003)   &     0.42    &     6$\pm$0.6      \\   
  21483   &   0.56    &     7            &    $\surd$   &    Wszolek \& Godlowski (2003)   &     0.56    &     7$\pm$0.7      \\   
  27778   &   0.35    &     5            &    $\surd$   &    Wszolek \& Godlowski (2003)   &     0.35    &     5$\pm$0.5      \\   
  34078   &   0.52    &     6            &    $\surd$   &    Wszolek \& Godlowski (2003)   &     0.52    &     6$\pm$0.6      \\   
  41117   &   0.45    &    10            &    $\surd$   &    Wszolek \& Godlowski (2003)   &     0.45    &    10$\pm$1.0      \\   
  42087   &   0.36    &     6            &    $\surd$   &    Wszolek \& Godlowski (2003)   &     0.36    &     6$\pm$0.6      \\   
  47129   &   0.34    &     3            &    $\times$  &    Wszolek \& Godlowski (2003)   &             &            \\   
 143275   &   0.17    &     1            &    $\times$  &    Wszolek \& Godlowski (2003)   &             &            \\
\enddata
\end{deluxetable}

\begin{deluxetable}{ccccccc}
\tabletypesize{\tiny}
\tablewidth{400pt}
\tablecaption{\footnotesize
              \label{tab:5844}
              Sources for the Equivalent Widths of the 5844$\Angstrom$ DIB.
              }
\tablehead{ \colhead{HD/BD} & \colhead{$E(B-V)$}  & \colhead{$W_{5844}$} & \colhead{Adopt"$\surd$"}  & \colhead{Sources} & \colhead{$E(B-V)$}  & \colhead{$W_{5844}$(m$\AngstromAL$)} 
\\
\colhead{} & \colhead{(mag)}  & \colhead{(m$\AngstromAL$)} & \colhead{or Reject"$\times$"}  & \colhead{}  & \colhead{This Work}   & \colhead{This Work}        
}
\startdata

 21291      &   0.42      &         40           &  $\surd$     &    Josafatsson \& Snow (1987)   &    0.42        &       40$\pm$4.0     \\ 
 30614      &   0.29      &         33           &  $\surd$     &    Josafatsson \& Snow (1987)   &    0.29        &       33$\pm$3.3     \\ 
 34078      &   0.52      &         35           &  $\surd$     &    Josafatsson \& Snow (1987)   &    0.52        &       35$\pm$3.5     \\ 
 34078      &   0.52      &        110           &  $\times$    &    Herbig (1975)                &                &              \\ 
 53974      &   0.31      &         32           &  $\surd$     &    Josafatsson \& Snow (1987)   &    0.31        &       32$\pm$3.2     \\ 
 147889     &   1.08      &        100           &  $\surd$     &    Herbig (1975)                &    1.08        &      100$\pm$10.0     \\ 
 147933     &   0.47      &         20           &  $\times$     &    Josafatsson \& Snow (1987)   &               &                       \\ 
 167971     &   1.08      &        137           &  $\surd$     &    Herbig (1975)                &    1.08        &      137$\pm$13.7     \\
\enddata
\end{deluxetable}

\begin{deluxetable}{ccccccc}
\tabletypesize{\tiny}
\tablewidth{400pt}
\tablecaption{\footnotesize
              \label{tab:5849}
              Sources for the Equivalent Widths of the 5849/5850$\Angstrom$ DIB.
              }
\tablehead{ \colhead{HD/BD} & \colhead{$E(B-V)$}  & \colhead{$W_{5849/5850}$} & \colhead{Adopt"$\surd$"}  & \colhead{Sources} & \colhead{$E(B-V)$}  & \colhead{$W_{5849/5850}$(m$\AngstromAL$)} 
\\
\colhead{} & \colhead{(mag)}  & \colhead{(m$\AngstromAL$)} & \colhead{or Reject"$\times$"}  & \colhead{}  & \colhead{This Work}   & \colhead{This Work}   
}
\startdata

  2905  &  0.33   &    20              &   $\surd$    &   Wszolek \& Godlowski (2003)      &         0.33    &   21.28$\pm$1.50         \\    
  2905  &  0.33   &    21.3$\pm0.3$    &   $\surd$    &   Megier et al.\ (2005)             &                 &                 \\             
  2905  &  0.32   &     87             &   $\times$   &   Herbig (1975)                    &                 &                 \\             
  2905  &  0.33   &    22              &   $\surd$    &   Herbig (1993)                    &                 &                 \\             
 15558  &  0.75   &    51.00$\pm3.75$  &   $\surd$    &   Desert et al.\ (1995)             &          0.75   &    51.00$\pm$3.75        \\    
 16691  &  0.83   &    29.05$\pm5.81$  &   $\surd$    &   Desert et al.\ (1995)             &          0.83   &    29.05$\pm$5.81        \\    
 21291  &  0.42   &    25              &   $\surd$    &   Wszolek \& Godlowski (2003)      &          0.41   &    24.40$\pm$1.90        \\    
 21291  &  0.42   &     38             &   $\times$   &   Josafasson \& Snow (1987)        &                 &                 \\
\enddata
\end{deluxetable}%


\end{document}